
\magnification=1200
\documentstyle{amsppt}
\language=0


\def\SLZ{\operatorname{SL}(2,\Z)}
\def\SLZp{\operatorname{SL}(2,\Z_{p})}
\def\SLZpl{\operatorname{SL}(2,\Z_{p^\lambda})}
\def\Spl#1#2{\operatorname{SL}(2,\Z_{#1^{#2}})}
\def\SL{\Gamma}
\def\GL#1{\operatorname{GL}(#1,\C)}
\def\GLK#1{\operatorname{GL}(#1,K)}
\def\Z{\Bbb Z}
\def\Q{\Bbb Q}
\def\R{\Bbb R}
\def\C{\Bbb C}
\def\N{\Bbb N}
\def\H{\Cal H}

\def\P{\frak P}
\def\min{\operatorname{min}}
\def\diag{\operatorname{diag}}
\def\trace{\operatorname{tr}}

\def\diag{\operatorname{diag}}
\def\w{{\Cal W}}
\def\id{{\text 1 \kern-2.8pt {\text I} }}
\def\mn{\medskip\smallskip\noindent}
\def\sn{\smallskip\smallskip\noindent}


\font\HUGE=cmbx12 scaled \magstep4
\font\Huge=cmbx10 scaled \magstep4
\font\Large=cmr12 scaled \magstep3

\font\large=cmr17 scaled \magstep0

\document
\nopagenumbers
\pageno = 0
\centerline{\HUGE Universit\"at Bonn}
\vskip 10pt
\centerline{\Huge Physikalisches Institut}
\vskip 2.5cm
\centerline{\Large On the Classification of}
\vskip 7pt
\centerline{\Large Modular Fusion Algebras}
\vskip 0.8cm
\centerline{\bf by}
\vskip 0.8cm
\centerline{\large Wolfgang Eholzer}
\vskip 1.2cm
\centerline{\bf Abstract}
\vskip3true mm
\noindent
We introduce the notion of (nondegenerate) strongly-modular fusion algebras.
Here strongly-modular means that the fusion algebra is induced via Verlinde's
formula by a representation of the modular group $\SL = \SLZ$ whose kernel
contains a congruence subgroup. Furthermore, nondegenerate means that
the conformal dimensions of possibly underlying rational conformal field
theories do not differ by integers.
Our main result is the classification of all strongly-modular fusion algebras
of dimension two, three and four and the classification of all nondegenerate
strongly-modular fusion algebras of dimension less than 24.
We use the classification of the irreducible representations of the
finite groups $\SLZpl$ where $p$ is a prime and $\lambda$ a positive integer.
Finally, we give polynomial realizations and fusion graphs for all simple
nondegenerate strong-modular fusion algebras of dimension less than 24.
\vskip 3pt
\noindent
\vskip 1.5cm
\settabs \+&  \hskip 110mm & \phantom{XXXXXXXXXXX} & \cr
\+ & Post address:                         & BONN--TH--94--18& \cr
\+ & Nu{\ss}allee 12                       & MPI-94-91       & \cr
\+ & D-53115 Bonn                          & hep-th/9408160  & \cr
\+ & Germany                               & Bonn University & \cr
\+ & e-mail:                               & August 1994     & \cr
\+ & eholzer\@mpim-bonn.mpg.de             &                 & \cr
\vfill
\eject


\def\Schellekens{1}
\def\MMS{2}
\def\Caselle{3}
\def\Ehof{4}
\def\Nobs{5}
\def\Fuchs{6}
\def\Duke{7}
\def\AMS{8}
\def\MIAU{9}
\def\huanglep{10}
\def\Zhu{11}
\def\Verlinde{12}
\def\Vafa{13}
\def\AndersonMoore{14}
\def\Gunnings{15}
\def\Dorn{16}
\def\BPZ{17}
\def\Wang{18}
\def\Don{19}
\def\Kir{20}
\def\GP{21}



\topmatter
	\toc
\head
1. Introduction
\endhead
\head
2. Rational conformal field theories and fusion algebras
\endhead
\subhead
2.1 Basic definitions
\endsubhead
\subhead
2.2 Some simple properties of modular fusion algebras
\endsubhead
\head
3. Some theorems on level $N$ representations of $\SLZ$
\endhead
\head
4. The classification of the irreducible level $p^\lambda$ representations
\endhead
\subhead
4.1 Weil representations associated to binary quadratic forms
\endsubhead
\subhead
4.2 The irreducible representations of $\SLZpl$ for $p\ne2$
\endsubhead
\subhead
4.2 The irreducible representations of $\Spl{2}{\lambda}$
\endsubhead
\head
5. Results on the classification of strongly-modular fusion algebras
\endhead
\subhead
5.1 Classification of the strongly-modular fusion algebras of dimension
    less than or equal to four
\endsubhead
\subhead
5.2 Proof of two Lemmas on diophantic equations
\endsubhead
\subhead
5.3 Classification of the nondegenerate strongly-modular fusion algebras
    of dimension less than 24
\endsubhead
\head
6. Conclusions
\endhead
\head
7. Appendix A: The irreducible level $p^\lambda$ representations
               of dimension less than or equal to four
\endhead
\head
8. Appendix B: Fusion matrices and graphs of the nondegenerate
               strongly-modular fusion algebras of dimension less than 24
\endhead
	\endtoc
\endtopmatter

\rightheadtext{On the classification of modular fusion algebras}
\leftheadtext{W\. Eholzer }

\document

\head
1. Introduction
\endhead

In the last ten years there have been several attempts for
the classification of rational conformal field theories (RCFTs).
However, a complete classification seems to be an impossible task
since, for example, all self dual double even lattices lead to RCFTs
and there is no hope to classify all such lattices of
rank greater than 24. Nevertheless, it might be possible to classify
all RCFTs with ``small'' effective central charge $\tilde c$.
The effective central charge is given by the difference of the central
charge and 24 times the smallest conformal dimension of the rational model
under consideration.
In particular, for $\tilde c \le 1$ a classification of RCFTs
can be obtained by using
a theorem of Serre-Stark describing all modular forms of weight $1/2$ on
congruence subgroups if one assumes that
the corresponding conformal characters are modular functions on
a congruence subgroup.

For $\tilde c>1$ only partial results have been obtained so far.
One of the possibilities is to look at RCFTs where the corresponding
fusion algebra has a ``small'' dimension.
In the special case of a trivial fusion algebra
the RCFT has only one superselection sector and
a classification of the corresponding modular invariant
partition functions for unitary theories with $c\le24$ has
been obtained \cite{\Schellekens}.
As a next step in the classification one can try to classify the
nontrivial fusion algebras of low dimension first and then
investigate corresponding RCFTs. Indeed, the modular fusion algebras of
dimension less than or equal to three satisfying the so-called
Fuchs conditions have been classified  (see e.g.\ \cite{\MMS,\Caselle}).
In this paper we develop several tools, following the ideas of
ref. \cite{\Ehof}, which enable us to classify
all strongly-modular fusion algebras of dimension less than or equal to four
(for a definition of strongly-modular fusion algebras see \S2).
Our approach is based on the known classification of the irreducible
representations of the groups $\Spl{p}{\lambda}$ \cite{\Nobs}.

Another possibility is to investigate theories where the corresponding
fusion algebra has a certain structure but may have arbitrary or
``big'' dimension.
Here, a classification of all selfconjugate fusion algebras which
are isomorphic to a polynomial ring in one variable where the distinguished
basis has a certain form and where the structure
constants are less than or equal to one has been obtained (see e.g.\
\cite{\Caselle}\footnotemark~).
Furthermore, a classification of all fusion algebras which are isomorphic
to a polynomial ring in one variable and where the quantum dimension
of the elementary field is smaller or equal to 2 is known
(this classification contains the fusion algebras occurring
in the classification of ref. \cite{\Caselle};
for a review  see e.g.\ \cite{\Fuchs}).
With the tools developed in this paper we obtain another partial
classification, namely of those strongly-modular fusion algebras
of dimension less than 24 where the corresponding representation $\rho$
of the modular group is such that $\rho(T)$ has nondegenerate eigenvalues.
The nondegeneracy of the eigenvalues of $\rho(T)$ means that the difference
of any two conformal dimensions of a possibly underlying RCFT is not
an integer.
The restriction on the dimension is of purely technical nature so that
it should  be possible to obtain a complete classification of all
nondegenerate strongly-modular fusion algebras with the methods
described in this paper by using systematically Galois theory.

\footnotetext{
More precisely, in \cite{\Caselle} all  selfconjugate modular fusion algebras
with $N_{i j}^k\le 1$, which are isomorphic to $\Q[x]/<P(x)>$ and
$\Phi_0 \cong 1, \Phi_1 \cong x, \Phi_j \cong p_j(x)\ (j=2,\dots,n-1)$
for some polynomials $P$ and $p_j$ and where the degree of $P$ is $n$
and the degree of the $p_j$ is $j$ have been classified
(the assumption on the degree of $p_j$ was used implicitly in loc.\ cit.\ ).}

This paper is organized as follows:
In \S 2 we recall some basic properties of rational conformal
field theories and give definitions of the relevant types of fusion algebras.
Section 3 contains some general theorems about representations
of the modular group which factor through a congruence subgroup.
In the next section we give a short review about the classification of
the irreducible representations of $\Spl{p}{\lambda}$ which will be
the main tool in the proof of the theorems in \S 5.
The main results of this paper are contained in \S 5. Here
we classify all strongly-modular fusion algebras of dimension less than or
equal
to four and the nondegenerate strongly-modular fusion algebras of dimension
less than 24. Finally, we summarize our results and point out some open
questions in the conclusion.
Two appendices contain the explicit form of the irreducible level
$p^\lambda$ representations of dimension less than or equal to four as well
as the fusion matrices and graphs of the simple nondegenerate
strongly-modular fusion algebras of dimension less than 24.

\head
2.  Rational conformal field theories and fusion algebras
\endhead

\mn
\subhead
2.1 Basic definitions
\endsubhead

Consider a chiral rational conformal field theory
(or rational model)  $R$ consisting of a symmetry algebra $\w$
and its finitely many inequivalent
irreducible modules $\H_i$ ($i=0,\dots,n-1$), i.e.\ $R$ is a rational
vertex operator algebra (RVOA) satisfying Zhu's finiteness condition
(for RVOA see e.g.\ \cite{\Duke,\AMS}  and for the connection of
RVOA to $\w$-algebras and rational models see  \cite{\MIAU}).
Here $\H_0$ denotes the vacuum representation.
For modules $\H$ of $\w$ there is the notion of conjugate (or adjoint or
dual) modules $\H'$. In particular, it is conjectured that one has
$(\H')' \cong \H$. Since $R$ is rational the conjugation defines a
permutation $\pi$ of order two of the irreducible modules
$\H'_i \cong \H_{\pi(i)}$.

The structure constants $N_{i,j}^k$ of the ``fusion algebra'' associated
to $R$ are given by the dimension of the corresponding space of intertwiners
of the modules $\H_i\otimes\H_j$ and $\H_k$ (for a definition
of intertwiners of modules of vertex operator algebras see e.g.\
\cite{\AMS}).
One of the important properties of the $N_{i,j}^k$ which is well
known in the physical literature is the fact that the numbers $N_{i,j}^k$
can be viewed  as the structure constants of an associative
commutative algebra, the fusion algebra. In the terminology of vertex
operator algebras a corresponding statement is proven under certain
assumptions in a recent series of papers \cite{\huanglep}.
In the abstract definition of fusion algebras the properties
of all known examples associated to RCFTs are collected.

\definition{Definition}
A {\bf fusion algebra} $\Cal F$ is a finite dimensional algebra over $\Q$
with a distinguished basis $\Phi_0=\id,\dots,\Phi_{n-1}$
($n=\dim(\Cal F )$) satisfying the following axioms:
\roster
\item
 $\Cal F$ is associative and commutative.
\item
 The structure constants $N_{i,j}^k$  ($i,j,k=0,\dots,n-1$)
 with respect to the distinguished basis $\Phi_i$ are nonnegative integers.
\item
 There exists a permutation $\pi\in S_n$ of order two such that for
 the structure constants in (2)
 one has
 $$ N_{i,j}^0 = \delta_{i,\pi(j)},\qquad
    N_{\pi(i),\pi(j)}^{\pi(k)} = N_{i,j}^k,\qquad  i,j,k=0,\dots,n-1.$$
\endroster
\enddefinition
\remark{Remarks}

An isomorphism $\phi$ of two fusion algebras ${\Cal F},{\Cal F}'$
is an isomorphism of unital algebras
which maps the distinguished basis to the distinguished basis, i.e.\
there exists a permutation $\sigma\in S_n$ such that
$\phi(\Phi_i) = \Phi'_{\sigma(i)}$ ($i=0,\dots, n-1$).

The tensor product of two fusion algebras $\Cal F$ and $\Cal F'$ is again
a fusion algebra, its distinguished basis is given by
$\Phi_{i_1}\otimes\Phi'_{i_2}$
($i_1=0,\dots,\dim(F)-1,\ i_2=0,\dots,\dim(F')-1$).

The permutation $\pi$ of order two is called charge conjugation.
Fusion algebras with trivial charge conjugation are called selfconjugate.

Note that it is an open question whether two nonisomorphic fusion
algebras can be isomorphic as unital algebras.
\endremark
\mn
It is known in many cases that fusion algebras arising from
RCFTs have additional properties. One of these additional
properties is their relation to conformal characters.
The conformal characters $\chi_i$ of the modules $\H_i$ of $\w$
are formal power series in $q$ defined by
$$ \chi_i(\tau) = \trace_{\H_i}(q^{L_0-\frac{c}{24}}) $$
where $L_0$ is the $0$-th Fourier mode of the chiral energy-momentum tensor
and $c$ is the central charge or the rank of the RVOA.
One can show for rational vertex operator algebras satisfying Zhu's
finiteness condition \cite{\Zhu} that the characters become holomorphic
functions  in the upper complex half plane by setting
$q = e^{2\pi i \tau}$. Furthermore, for these RVOAs the space spanned by
the finitely many conformal characters  is invariant under the action
of the modular group $\SL = \SLZ$.
Indeed, it is conjectured that Zhu's finiteness condition is not
necessary at all.
It was conjectured in 1988 by E.\ Verlinde \cite{\Verlinde} that
for any rational model there exists a
representation  $\rho:\SL\to\GL{n}$ of $\SL$ such that
$$\align
  \chi_i(A\tau) &= (\chi_i|A)(\tau)
               = \sum_{m=0}^{n-1} \rho(A)_{j,i} \chi_j(\tau)
  \qquad A \in \SL \\
  N_{i,j}^0 &=  \rho(S^2)_{i,j} \\
  N_{i,j}^k &= \sum_{m=0}^{n-1}
               \frac{\rho(S)_{i,m}\rho(S)_{j,m}\rho(S^{-1})_{m,k}}
                    {\rho(S)_{0,m}}. \\
\endalign$$
We will refer to this formula as ``Verlinde's formula'' in the
following.
The above conjecture motivates the definition of modular fusion algebras.

\definition{Definition}
A {\bf modular fusion algebra} $({\Cal F},\rho)$ is a fusion algebra
$\Cal F$ together with a unitary representation $\rho:\SLZ\to\GL{n}$
satisfying the following additional axioms:
\roster
\item
 $\rho(S)$ is a symmetric and $\rho(T)$ is a diagonal matrix.
\item
 $N_{i,j}^0 = \rho(S^2)_{i,j}$,
\item
 $N_{i,j}^k = \sum_{m=0}^{n-1}
               \frac{\rho(S)_{i,m}\rho(S)_{j,m}\rho(S^{-1})_{m,k}}
                    {\rho(S)_{0,m}}$
\endroster
where $N_{i,j}^k$ ($i,j,k=0,\dots,n-1$) are the structure constants of $\Cal F$
with respect to the distinguished basis.
\enddefinition
\remark{Remarks}

Note that property (3) already implies that $\Cal F$ is associative
and commutative.

Two modular fusion algebras $({\Cal F},\rho)$ and $({\Cal F'},\rho')$ are
called isomorphic
if: 1) $\Cal F$ and $\Cal F'$ are isomorphic as fusion algebras,
    2) $\rho$ and $\rho'$ are equivalent,
    3) $\rho(T)_{i,j} = \rho'(T)_{\sigma(i) \sigma(j)}$ where
       $\sigma\in S_n$ is the permutation defined by the isomorphism
       of the fusion algebras.

The tensor product of two modular fusion algebras
$({\Cal F},\rho),({\Cal F'},\rho')$ is defined by
$({\Cal F}\otimes {\Cal F'},\rho\otimes\rho')$ and is
again a  modular fusion algebra.

A (modular) fusion algebra is called composite if it is isomorphic to
a tensor products of two nontrivial (modular) fusion algebras.
Here a (modular) fusion algebra is called trivial if it is one dimensional. A
noncomposite (modular) fusion algebra is also
called simple.

Note that for a modular fusion algebra with trivial charge conjugation
($\rho(S^2) = \id$) the matrix $\rho(S)$ is real.

For modular fusion algebras associated to rational models
the eigenvalues of $\rho(T)$ are given by
the conformal dimensions $h_i$ ($i=0,\dots,n-1$) of the irreducible
modules $\H_i$ ($h_i$ is the smallest $L_0$ eigenvalue in the module
$\H_i$) and the central charge $c$ of the theory:
$$\rho(T) = \diag(e^{2\pi i (h_0-c/24)},\dots,e^{2\pi i (h_{n-1}-c/24})).$$

Quite often nonisomorphic modular fusion algebras are isomorphic as
fusion algebras.
\endremark

In the later sections we will investigate which representations of
$\SL$ are related to modular fusion algebras.

\definition{Definition}
A representation $\rho:\SLZ\to\GL{n}$ of the modular group is called
{\bf conformally admissible} or simply {\bf admissible} if there
exists a fusion algebra $\Cal F$ such that $({\Cal F},\rho)$ is a
modular fusion algebra.
\enddefinition

It is known that modular fusion algebras associated to rational
models have many additional properties.
In particular, the central charge and the conformal dimensions
are rational \cite{\Vafa,\AndersonMoore}.
Furthermore, compatibility conditions between the central charge $c$, the
conformal dimensions $h_i$ and the fusion coefficients $N_{i j}^k$
(the so-called Fuchs conditions) are satisfied
(see e.g.\ \cite{\MMS} or \cite{\Caselle}\footnotemark):

\footnotetext{ Note that the formula connecting the central
 charge with the conformal dimension in \cite{\Caselle}
 contains a misprint.}

$$ \align
   & \frac{n(n-1)}{12}
     -\sum_{m=0}^{n-1} \left( h_i- \frac{c}{24} \right)
     \in \frac16 (\N \backslash \{1\}), \\
    &\sum_{m=0}^{n-1}
    \left(
    (h_i+h_j+h_k+h_l) N_{i,j}^m N_{k,m}^l
    -h_m (N_{i,j}^m N_{k,m}^l + N_{i,k}^m N_{j,m}^l + N_{i,l}^m N_{k,j}^m)
    \right) \\
   &-\frac12 \left( \sum_{m=0}^{n-1} N_{i,j}^m N_{k,m}^l \right)
           \left( 1- \sum_{m=0}^{n-1} N_{i,j}^m N_{k,m}^l \right) \in \N
\endalign$$
However, in this paper we will not make any use of these properties.

Instead we will extensively rely on the observation that
in all known examples of RCFTs the conformal characters are modular
functions on some congruence subgroup of $\SL$. Therefore, the
corresponding representation $\rho$ factors through a representation
of $\Gamma_N$. Here we have used $\Gamma_N$ for
the principal congruence subgroup of $\SL$ of level $N$
$$ \Gamma_N = \{\ A\in\SL\ \vert\ A\equiv\id \bmod N\ \}. $$

\definition{Definition}
A modular fusion algebra $({\Cal F},\rho)$ is called {\bf strongly-modular}
if the kernel of the representation $\rho$ contains a congruence
subgroup of $\SL$.
\enddefinition

In this case $\rho$ defines a representation of $\Spl{N}{}$  and is
called a level $N$ representation of $\SL$
(here and in the following we use $\Z_N$ for $\Z/N\Z$).
A level $N$ representation $\rho$ will be called even or odd
if $\rho(S^2) = \id$ or $\rho(S^2)=-\id$, respectively.
Furthermore, one can show that for strongly-modular fusion algebras
associated to rational models the representation
$\rho$ is defined over the field $K$ of $N$-th roots of unity, i.e.\
$\rho:\SL\to\GLK{n}$ if the corresponding conformal characters are modular
functions on some congruence subgroup \cite{\MIAU}.
Indeed, we expect that this is true for all RCFTs what motivates the
following definition and conjecture.

\definition{Definition}
A level $N$ representation $\rho:\SLZ \to \GL{n}$ is called
$K$-rational if it is defined over the field $K$ of the $N$-th roots of
unity, i.e.\ $\rho:\SLZ \to \GLK{n}$.
\enddefinition

\definition{Conjecture}
All modular fusion algebras associated to rational models
are strongly-modular fusion algebras and
the corresponding representations of the modular group are $K$-rational.
\enddefinition

\subhead
2.2 Some simple properties of modular fusion algebras
\endsubhead

In this section we prove some simple lemmas about modular fusion
algebras which will be needed in the proofs of the main theorems
in \S 5.

\proclaim{Lemma 1}
Let $({\Cal F},\rho)$ be a modular fusion algebra.
Assume that $\rho(T)$ has nondegenerate eigenvalues.
Then $\rho$ is irreducible.
\endproclaim
\demo{Proof}
 Assume that $\rho$ is reducible and $\rho(T)$ has nondegenerate
 eigenvalues. Then $\rho(S)$ has block diagonal form and
 therefore $\rho(S)_{0,m}=0$ for some $m$.
 This is a contradiction to property (3) in the
 definition of modular fusion algebras.
\enddemo

\definition{Definition}
A modular fusion algebra $({\Cal F},\rho)$ is called {\bf degenerate} or
{\bf nondegenerate} if $\rho(T)$ has degenerate or nondegenerate
eigenvalues, respectively.
\enddefinition

\proclaim{Lemma 2}
Let $\rho,\rho':\SL\to\GL{n}$ be equivalent, irreducible, unitary
representations of the modular group.
Assume that $\rho(T) = \rho'(T)$ is a diagonal matrix with
nondegenerate eigenvalues.
Then there exists a unitary diagonal matrix $D$ such that
$\rho = D^{-1}\rho' D$.
\endproclaim
\demo{Proof}
Since $\rho$ and $\rho'$ are equivalent there exists a matrix $D'$ such that
$\rho = {D'}^{-1}\rho' D'$. Since $\rho(T) = \rho'(T)$ is a diagonal
matrix with nondegenerate eigenvalues $D'$ is diagonal.
Finally, the irreducibility of
$\rho$ implies by Schur's lemma that ${D'}^+ D' = \alpha\id$ for some positive
real number $\alpha$ so that $D=\frac1{\sqrt\alpha}D'$ satisfies the
desired  properties.
\enddemo

\proclaim{Lemma 3}
Let $({\Cal F},\rho)$ and $({\Cal F'},\rho')$ be two nondegenerate modular
fusion algebras.
Assume that $\rho$ is equivalent to $\rho'$ and $\rho(T) = \rho'(T)$.
Then $\Cal F$ and $\Cal F'$ are isomorphic as fusion algebras.
\endproclaim
\demo{Proof}
The lemma follows directly from the definition of (modular) fusion
algebras and Lemma 2.
\enddemo

\proclaim{Lemma 4}
Let $({\Cal F},\rho)$ be a modular fusion algebra.
Then $\rho$ is not isomorphic to a direct sum of one
dimensional representations.
\endproclaim
\demo{Proof}
If $\rho$ is the direct sum of one dimensional  representations
$\rho(S)$ is also a diagonal matrix. This implies that one cannot apply
Verlinde's formula giving a contradiction since we have assumed
that $({\Cal F},\rho)$ is a modular fusion algebra.
\enddemo

Since there are exactly 12 one dimensional representations
of $\SL$ one has the following trivial lemma.
\proclaim{Lemma 5}
\roster
\item
Let $\rho$ be a one dimensional representation of $\SL$.
Then $\rho$ is equivalent to one of the following representations
$$ \rho(S) = e^{2\pi i \frac{3n}4},\qquad \rho(T) = e^{2\pi i \frac{n}{12}},
   \qquad n=0,\dots,11.$$
\item
Let $({\Cal F},\rho)$ be a one dimensional modular fusion algebra.
Then $({\Cal F},\rho)$ is strongly-modular, $\Cal F$ is trivial and $\rho$ is
given by
$$ \rho(S) = (-1)^n,\qquad
   \rho(T) = e^{2\pi i \frac{n}6},\qquad n=0,\dots,5. $$
\endroster
\endproclaim

\proclaim{Lemma 6}
Let $({\Cal F},\rho)$ be a strongly-modular fusion algebra associated to a
rational model. Then $\rho$ is $K$-rational.
\endproclaim
\demo{Proof}
For a rational vertex operator algebra satisfying Zhu's finiteness condition
the characters are holomorphic functions on the upper complex half plane.
Since we have assumed that $({\Cal F},\rho)$ is strongly-modular $\rho$ is a
level $N$ representation for some $N$. This implies that the characters
are modular functions on $\Gamma_N$. Moreover, their Fourier coefficients
are positive integer so that one can apply the theorem on $K$-rationality
of ref. \cite{\MIAU} implying that $\rho$ is $K$-rational.
\enddemo

Although Lemma 6 will not be used in the following it provides us with a good
motivation for looking at $K$-rationality of level $N$
representations.

\head
3. Some theorems on level $N$ representations of $\SL$
\endhead

In this section we will consider level $N$ representations of $\SLZ$.
Firstly, we review that all irreducible representations of
$\Spl{N}{}$ can be obtained by those of $\Spl{p}{\lambda}$
where $p$ is a prime and $\lambda$ is a positive integer.
Secondly, we discuss the construction of level $p^\lambda$
representations using Weil representations (in this part we follow
ref. \cite{\Nobs}).

\proclaim{Lemma 7}
Let $\rho$ be a finite dimensional representation of $\Spl{N}{}$
where $N$ is a positive integer.
Then the representation $\rho$ is completely reducible.
Furthermore, each irreducible component $\omega$ of $\rho$
has a unique product decomposition
$$ \omega \cong \otimes_{j=1}^{n} \pi({p_j^{\lambda_j}}) $$
where $N=\prod_{j=1}^{n} p_j^{\lambda_j}$ is the prime factor
decomposition of $N$ and the $\pi({p_j^{\lambda_j}})$ are
irreducible representations of $\Spl{p_j}{\lambda_j}$.
\endproclaim
\demo{Proof}
Since $\Spl{N}{}$ is a finite group $\rho$ is completely reducible.
The second statement, namely that the irreducible representations of
$\Spl{N}{}$ can be written as a tensor product of irreducible
representations of $\Spl{p_j}{\lambda_j}$ where
$N=\prod_{j=1}^{n} p_j^{\lambda_j}$ is the prime factor
decomposition of $N$, can be seen as follows.
For a proof of the second statement note that
$$\Spl{N}{} = \Spl{p_1}{\lambda_1}\times\dots\times\Spl{p_n}{\lambda_n}$$
where $N=\prod_{j=1}^{n} p_j^{\lambda_j}$ (see e.g.\ \cite{\Gunnings}).
Obviously, the tensor product of irreducible representations
$\pi({p_j^{\lambda_j}})$ of $\Spl{p_j}{\lambda_j}$ is an
irreducible representation of $\Spl{N}{}$. Using now
Burnside's lemma we obtain the second statement.
\enddemo

In order to deal with the representations of the groups
$\Spl{p}{\lambda}$ we describe their structure by the following theorem.

\proclaim{Theorem 1 \cite{\Nobs, Satz 1, p. 466}}
The group $\Spl{p}{\lambda}$ is generated by the elements
$$ S= \left(\matrix 0&-1\\1&0\endmatrix\right),\quad
   T= \left(\matrix 1&1\\0&1\endmatrix\right),\quad
$$
and the relations
$$\align
   T^{p^\lambda} & = \id,\qquad
   S^2 = H(-1) \\
   H(a) H(a') &= H(a a'), \qquad
   H(a) T     = T^{a^2} H(a),\qquad
   S H(a)     = H(a^{-1}) S   \\
\endalign $$
where  $ H(a) := T^{-a} S T^{-a^{-1}} S^{-1} T^{-a} S^{-1}$
and  $a,a' \in \Z_{p^\lambda}^*$.
\endproclaim
\remark{Remark}
As elements of $\Spl{p}{\lambda}$ the  $H(a)$
($a\in \Z_{p^\lambda}^*$) are given by
$$ H(a) = \left(\matrix a&0\\0&a^{-1}\endmatrix\right).$$
\endremark

We will now describe the construction of representations of $\Spl{p}{\lambda}$
by means of Weil representations.

\definition{Definition}
Let $M$ be a finite $\Z_{p^\lambda}$ module. A quadratic form $Q$ of $M$
is a map $Q:M\to\ p^{-\lambda}\Z / \Z$ such that
\roster
\item
 $Q(-x) = Q(x)$ for all $x\in M$.
\item
 $B(x,y) := Q(x+y) - Q(x) -Q(y)$ defines a  $\Z_{p^\lambda}$-bilinear map
 from $M\times M$ to $  p^{-\lambda}\Z / \Z$.
\endroster
\enddefinition

\definition{Definition}
A finite $\Z_{p^\lambda}$ module $M$ together with a quadratic form $Q$
is called a quadratic module of $\Z_{p^\lambda}$.
\enddefinition

\definition{Definition}
Let $(M,Q)$ be a quadratic module.
Define a right action of $\Spl{p}{\lambda}$ on the space of $\C$ valued
functions on $M$ by
$$ \align
  (f|T)(x)       &= e^{2\pi i Q(x)} \ f(x) \\
  (f|H(a))(x)    &= \alpha_Q(a) \alpha_Q(-1) \ f(x)
                    \qquad \forall a \in \Z_{p^\lambda}^* \\
  (f|S^{-1})(x)  &= \frac{\alpha_Q(-1)}{\vert M\vert ^{1/2}}
                       \sum_{y\in M}  e^{2\pi i B(x,y)} \ f(y) \\
\endalign
$$
where $ \vert M \vert$ denotes the order of $M$,
$$ \alpha_Q(a) = \frac1{\vert M\vert} \sum_{x\in M} e^{2\pi i a Q(x)} $$
and $f$ is any $\C$ valued function on $M$.

If this right action of $\Spl{p}{\lambda}$ defines a representation of
$\Spl{p}{\lambda}$ it is called  the Weil representation associated to the
quadratic module $(M,Q)$ and denoted by $W(M,Q)$.
\enddefinition

Note that the above right action always defines a projective
representation of $\SL$. A necessary and sufficient condition for it
to define a proper representation is given by the following theorem.
\proclaim{Theorem 2 \cite{\Nobs, Satz 2, p. 467} }
The above right action of $\Spl{p}{\lambda}$ defines a representation of
$\Spl{p}{\lambda}$ if and only if
$$ \alpha_Q(a) \alpha_Q(a') = \alpha_Q(1) \alpha_Q(a a') \
   \qquad a,a'\in \Z_{p^\lambda}^*. $$
\endproclaim

\head
4. The classification of the irreducible level $p^\lambda$ representations
\endhead

Although the classification of the irreducible representations of
the finite groups $\Spl{p}{\lambda}$ is contained in \cite{\Nobs}
we will give a short review here. Our main motivation for this is the
fact that we will strongly rely on this classification in the proofs
of the main theorems in \S 5. Furthermore, ref. \cite{\Nobs} is not
written in English but in German.

In the first subsection we describe how one can obtain
irreducible level $p^\lambda$ representations as
subrepresentations of Weil representations.
The second and third subsection are used to give complete lists of the
corresponding representations for the cases of $p\not=2$ and $p=2$,
respectively.

In addition to the review we investigate in some cases
whether the irreducible representations are $K$-rational or not.

\subhead
4.1 Weil representations associated to binary quadratic forms
\endsubhead

Most of the irreducible representations of $\Spl{p}{\lambda}$ can be obtained
as subrepresentations of Weil representations $W(M,Q)$ associated to
a module $M$ of rank one or two.
The following two theorems describe the Weil representations needed in the
later sections.

\proclaim{Theorem 3 \cite{\Nobs, Lemma 1, Satz 3, p.\ 474} }
Let $p\ne 2 $ be a prime. Then the following quadratic modules of
$\Z_{p^\lambda}$ define Weil representations:
$$ \align
  &(1)\quad M = \Z_{p^\lambda},\qquad \qquad  \ \ \,
            Q(x) = p^{-\lambda} r x^2
      \qquad \qquad \quad \quad\, (\lambda \ge 1)
      \qquad \left( R_{\lambda}(r) \right) \\
  &(2)\quad M = \Z_{p^\lambda}\oplus\Z_{p^\lambda},\quad \quad\,
            Q(x) = p^{-\lambda}x_1 x_2
        \qquad \qquad\quad\ \, \,(\lambda \ge 1)
        \qquad \left( D_{\lambda} \right) \\
  &(3)\quad M = \Z_{p^\lambda}\oplus\Z_{p^\lambda},\quad \quad\,
            Q(x) = p^{-\lambda}(x_1^2 -u x_2^2)
        \qquad \quad\, (\lambda \ge 1)
        \qquad \left( N_{\lambda} \right) \\
  &(4)\quad M = \Z_{p^\lambda}\oplus\Z_{p^{\lambda-\sigma}},\quad \,
            Q(x) = p^{-\lambda} r (x_1^2 -p^\sigma t x_2^2)
        \qquad (\lambda \ge 2)
        \qquad \left( R_\lambda^\sigma (r,t) \right)
\endalign $$
where $r,t$ run through $\{1,u \}$ with $(\frac u p) = -1$
($(\frac{\cdot}{\cdot})$ denotes the Legendre symbol),
where $\sigma = 1,\dots,\lambda-1$ and where the last column contains
the name of the corresponding Weil representation.
\endproclaim

\proclaim{Theorem 4 \cite{\Nobs, Satz 4, p.\ 474} }
Let $p=2$. Then the following quadratic modules of $\Z_{2^\lambda}$
define Weil representations:
$$ \align
  &(1)\quad M = \Z_{2^\lambda}\oplus\Z_{2^\lambda},\qquad \quad\ \
            Q(x) = 2^{-\lambda}x_1 x_2
            \qquad \qquad \quad \ \,(\lambda \ge 1)
            \qquad \left( D_{\lambda} \right)  \\
  &(2)\quad M = \Z_{2^\lambda}\oplus\Z_{2^\lambda},\qquad \quad\ \
            Q(x) = 2^{-\lambda}(x_1^2 + x_1 x_2 + x_2^2)
            \ (\lambda \ge 1)
            \qquad \left( N_{\lambda} \right) \\
  &(3)\quad M = \Z_{2^{\lambda-1}}\oplus\Z_{2^{\lambda-\sigma-1}},\quad
            Q(x) = 2^{-\lambda} r (x_1^2 +2^\sigma t x_2^2)
        \quad\ \ (\lambda \ge 2)
            \qquad \left( R_\lambda^\sigma (r,t) \right)
\endalign $$
where $\sigma = 0,\dots,\lambda-2$,
where $(r,t)$ run through a system of representatives of the classes
of pairs defined by $ (r_1,t_1) \cong (r_2,t_2)$
\sn
if $t_1 \equiv t_2 \bmod \min(8,2^{\lambda-\sigma})$
and
$$\cases
     r_2 \equiv r_1 \bmod 4 \quad \text{or} \quad
     r_2 \equiv r_1 t_1 \bmod 4   &\text{for} \quad \sigma=0\\
     r_2 \equiv r_1 \bmod 8 \quad \text{or} \quad
     r_2 \equiv r_1 + 2r_1 t_1 \bmod 8 &\text{for} \quad \sigma=1\\
     r_2 \equiv r_1 \bmod 4  &\text{for} \quad \sigma=2\\
     r_2 \equiv r_1 \bmod 8  &\text{for} \quad \sigma\ge3\\
\endcases$$
and where the last column contains the name of the corresponding
Weil representation.
\endproclaim

All irreducible representations of $\Spl{p}{\lambda}$ can be obtained
as subrepresentations of Weil representations $W(M,Q)$.
One possibility to extract subrepresentations of such representations
is to use characters of the automorphism group of the quadratic form
$Q$:
\mn
\proclaim{Theorem 5 (see e.g.\ \cite{\Nobs}) }
Let $W(M,Q)$ be a Weil representation described by Theorem 3 or 4,
$\Cal U$ an abelian subgroup of $\text{Aut}(M,Q)$ and $\chi$ a
character of $\Cal U$.
Then the subspace
$$V(\chi) := \{ \ f:M\to \C\  \vert\  f(\epsilon x) = \chi(\epsilon) f(x),
               \quad x\in M,\ \epsilon \in {\Cal U} \ \}$$
of $\C^M$ is invariant under $\Spl{p}{\lambda}$. The corresponding
subrepresentation is denoted by $W(M,Q,\chi)$.
\endproclaim
\remark{Remarks}

(a) The space $V(\chi)$ is spanned by $ V(\chi) = < f_x(\chi)>_{x\in M} $
where
$$f_x(\chi)(y) = \sum_{\epsilon\in {\Cal U}} \chi(\epsilon)
                                              \delta_{\epsilon x,y},
  \qquad  \delta_{x,y} = \cases 1 & \text{for}\ $x=y$ \\
                                0 & \text{otherwise.}
                        \endcases
$$

(b) The automorphism group of the quadratic forms in Theorem 4 contain a
conjugation $\kappa$: $\kappa(x_1,x_2) = (x_2,x_1)$ in case (1) and
 $\kappa(x_1,x_2) = (x_1,-x_2)$  in the cases (2) and (3).
In these cases the space
$$V(\chi)_{\pm} := \{ \ f\in V(\chi)\ \vert\  f(\kappa x) = \pm f(x),
                     \quad x\in M \ \}$$
is invariant under $\Spl{2}{\lambda}$.  The corresponding
subrepresentation is denoted by $W(M,Q,\chi)_\pm$.
\endremark

{}From now on we will denote the trivial character $\chi \equiv 1$ by
$\chi_1$. Indeed, almost all irreducible representations of $\Spl{p}{\lambda}$
can be obtained as subrepresentations of the Weil representations described
by Theorems 3 and 4 using ``primitive'' characters:

\definition{Definition\footnotemark}
Let $W(M,Q)$ be a Weil representation described by Theorem 3 or 4
and let ${\Cal U} = \text{Aut}(M,Q)$. A character $\chi$ of $\Cal U$ is called
primitive iff there exists an element $\epsilon\in {\Cal U}$ with
$\chi(\epsilon)\not=1$ such that each element of $pM$ is a fixed point of
$\epsilon$. The set of primitive characters of $\Cal U$ will be denoted by
$\P$.
\enddefinition

\footnotetext{ In the case of $M=\Z_{2^{\lambda-1}}\oplus\Z_{2}$
 ($\lambda \ge 5$) the definition of primitive characters is
 slightly different \cite{\Nobs, p.\ 491}:
 Here $\Cal U \cong <-1> <\alpha>$ with
 $\alpha = \cases 1+4t+\sqrt{-8t} & \lambda=5\\
                  1-2^{\lambda-3}+\sqrt{-2^{\lambda-2}t} & \lambda>5
           \endcases\qquad\qquad\qquad\qquad$
 and $\chi$ is primitive if $\chi(\alpha) = -1$. }

With this definition we have:
\proclaim{Theorem 6 \cite{\Nobs, Hauptsatz 1, p. 492}}
Let $W(M,Q)$ and $W(M',Q')$ be Weil representation described by
Theorem 3 or 4 and  $\chi,\chi'$ primitive characters. Then one has
\roster
\item
 $W(M,Q,\chi)$ is an irreducible level $p^\lambda$ representation.
\item
 $W(M,Q,\chi)$ and $W(M',Q',\chi')$ are isomorphic if and only if
 the quadratic modules $(M,Q)$ and $(M',Q')$ are isomorphic and
 $\chi = \chi'$ or $\chi= \bar\chi'$.
\endroster
\endproclaim

The second main theorem of ref. \cite{\Nobs} describes
the classification of the
irreducible representations of $\Spl{p}{\lambda}$.

\proclaim{ Theorem 7 \cite{\Nobs, Hauptsatz 2, p.\ 493}}
The Weil representations described by the Theorems 3 and 4 contain all
irreducible representations of the groups $\Spl{p}{\lambda}$
(in general they are of the form $W(M,Q,\chi)$ for a primitive character
$\chi$) apart from 18 exceptional representations for $p=2$. Theses exceptional
representations can be obtained as tensor products of two representations
contained in some $W(M,Q)$ (described by Theorem 3 or 4).
\endproclaim

Complete lists of irreducible representations of $\Spl{p}{\lambda}$
will be given in \S 4.2 and \S 4.3.

\subhead
4.2 The irreducible representations of $\SLZpl$ for $p\not=2$
\endsubhead

In the classification of the irreducible representations of
$\SLZpl$ for $p\not=2$ one has to distinguish the cases $\lambda=1$
and $\lambda>1$. Therefore, we treat them separately.

Following \cite{\Nobs} we denote the trivial representation by $C_1$.

\proclaim{Theorem 8 \cite{\Nobs}}
A complete set of irreducible representations of $\SLZp$ for
a prime $p$ with $p\not= 2$ is given by the representations collected in
Table~1. In Table 1 the $\chi$ run through the set of characters of $\Cal U$
and $\chi_{-1}$ is the unique nontrivial character of $\Cal U$
taking values in $\pm1$. Furthermore, we denote by $\#$ (here
and in the following) the number of inequivalent representations.
\endproclaim
\mn
\centerline{Table 1: Irreducible representations of $\SLZp$ for $p\not=2$}
\smallskip\noindent
\centerline{
\vbox{ \offinterlineskip
\def\Tablespace{ height2pt&\omit&&\omit&&\omit&&\omit&\cr }
\def\Tablerule{ \Tablespace
                \noalign{\hrule}
                \Tablespace      }
\hrule
\halign{&\vrule#&
  \strut\quad\hfil#\hfil\quad\cr
\Tablespace
& type of rep. &&  && dimension  && \# &\cr
\Tablerule
& $D_1(\chi)$        && $\chi\in\P$ && $p+1$
  && $\frac12(p-3)$ &\cr \Tablerule
& $N_1(\chi)$        && $\chi\in\P$ && $p-1$
  && $\frac12(p-1)$ &\cr \Tablerule
& $R_1(r,\chi_1)$    && $\bigl({r\over p} \bigr) = \pm 1 $
  && $\frac12(p+1)$   && $2$            &\cr \Tablerule
& $R_1(r,\chi_{-1})$ && $\bigl({r\over p} \bigr) = \pm 1$
  && $\frac12(p-1)$  && $2$            &\cr \Tablerule
& $N_1(\chi_1)$      && \omit
  && $p$            && $1$            &\cr \Tablespace
}
\hrule}
}
\mn

We will denote the 3 one dimensional level 3 representations
$C_1$, $R_1(1,\chi_{-1})$ and $R_1(2,\chi_{-1})$ by $B_1$, $B_2$ and
$B_3$, respectively.
\smallskip
The explicit form of these representations is well known
(see e.g.\ \cite{\Ehof})
and one can address the question which of these representations are
$K$-rational (see also \cite{\MIAU}). Note that, in view of the results
in \S 2.2,
this question is natural in the context of admissible representations.
\proclaim{Lemma 8}
Let $p\not=2$ be a prime.
\roster
\item
For $p\equiv 1 \pmod 3$ there is exactly one and for
$p \not\equiv 1 \pmod 3$ there is no $K$-rational representation of
type $D_1(\chi)$.
\item
For $p\equiv 2 \pmod 3$ there is exactly one and for
$p \not\equiv 2 \pmod 3$ there is no $K$-rational
representation of type $N_1(\chi)$ ($\chi\in\P$).
\item
The representations of type $R_1(r,\chi_{\pm 1})$ and
$N_1(\chi_1)$ are $K$-rational.
\endroster
\endproclaim
\demo{Proof}
Using a character table for the above representations (see e.g.\ \cite{\Dorn})
one easily finds that the characters of representations of type
$D_1(\chi)$ or $N_1(\chi)$ take values in the field of $p$-th roots
of unity only if $p\equiv 1 \pmod 3$ or $p\equiv 2 \pmod 3$
and if $\chi$ is a character of order 3.
Therefore, there is at most one $K$-rational representation of
type $D_1(\chi)$ or $N_1(\chi)$ for the corresponding values of $p$.
Using the explicit form of these representations  (see e.g.\ \cite{\Ehof})
one finds that these two representations are indeed $K$-rational.
For the other two types of representations the $K$-rationality follows
directly from the fact that $\chi_{\pm 1}$ takes values in $\pm1$.
\enddemo

\proclaim{Theorem 9 \cite{\Nobs}}
A complete set of irreducible representations of $\Spl{p}{\lambda}$
for $p\not=2$ prime and $\lambda>1$ is given by the representations
in Table 2.
Where $\chi_{-1}$ is the unique nontrivial
character with values in $\pm1$ and $R_\lambda(r,\chi_{\pm1})_1$ is the unique
level $p^\lambda$ subrepresentation of $R_\lambda(r,\chi_{\pm1})$
which has dimension $\frac12(p^2-1)p^{\lambda-2}$.
\endproclaim

\eject
\centerline{Table 2: Irreducible representations of $\SLZpl$ for
                     $p\not=2$ and $\lambda>1$ }
\smallskip\noindent
\centerline{
\vbox{ \offinterlineskip
\def\Tablespace{ height2pt&\omit&&\omit&&\omit&&\omit&\cr }
\def\Tablerule{ \Tablespace
                \noalign{\hrule}
                \Tablespace      }
\hrule
\halign{&\vrule#&
  \strut\quad\hfil#\hfil\quad\cr
\Tablespace
& type of rep. &&  && dimension  && \# &\cr
\Tablerule
& $D_\lambda(\chi)$    && $\chi\in \P$ && $(p+1)p^{\lambda-1}$
  && $\frac12(p-1)^2p^{\lambda-2}$                             &\cr \Tablerule
& $N_\lambda(\chi)$    && $\chi\in \P$ && $(p-1)p^{\lambda-1}$
  && $\frac12(p^2-1)p^{\lambda-2}$                             &\cr \Tablerule
& $R_\lambda^\sigma(r,t,\chi)$
  && $\bigl({r\over p}\bigr) = \pm 1,\ \bigl({t\over p}\bigr) = \pm 1$
  && $\frac12(p^2-1)p^{\lambda-2}$
  && $4 \sum_{\sigma=1}^{\lambda-1} (p-1)p^{\lambda-\sigma-1}$ &\cr \Tablerule
& $R_\lambda(r,\chi_{\pm1})_1$ && $\bigl({r\over p}\bigr) = \pm 1$
  && $\frac12(p^2-1)p^{\lambda-2}$ && $4$                      &\cr \Tablespace
}
\hrule}
}

\proclaim{Lemma 9}
Let $p\not=2$ be a prime and $\lambda>1$ an integer.
\roster
\item
The representations of type  $R_\lambda^\sigma(r,t,\chi)$
are $K$-rational for $p\not=2$ and $\lambda>1$.
\item
The representations of type $R_\lambda(r,\chi_{\pm1})_1$
are $K$-rational for $p\not=2$ and $\lambda>1$.
Furthermore, the image of $T$ under these representations
has nondegenerate eigenvalues only if  $p=3$ and $\lambda=2$.
\endroster
\endproclaim
\demo{Proof}
Since the automorphism group of the quadratic form of
$R_\lambda^\sigma(r,t,\chi)$ is given by \cite{\Nobs, p. 495}
${\Cal U}  \cong \Z_2 \times \Z_{p^{\lambda-\sigma}}$
we obtain (1).
In the second case one obviously has ${\Cal U} \cong \Z_2$ so that
the $K$-rationality follows directly.
The statement concerning the eigenvalues of the image of $T$ for the
representations of type $R_\lambda(r,\chi_{\pm1})_1$ is proved in
Satz 4 of \cite{\Nobs}.
\enddemo

\subhead
4.3 The irreducible representations of $\Spl{2}{\lambda}$
\endsubhead

The classification of the irreducible representations
of $\Spl{2}{\lambda}$ is complicated since there are a lot of
exceptional representations for $\lambda < 6$ \cite{\Nobs}.
Since these representations have small dimensions
and we will be interested in such representations
in \S 5 we describe them in the rest of this subsection.
The Tables 3-8 list complete sets of irreducible representations of the groups
$\Spl{2}{\lambda}$ for the corresponding values of $\lambda$.

For $\lambda=1$ there are only two irreducible representations (see Table 3).
The representation $C_2$ is given by $C_2(S) = C_2(T) = -1$ and
both level 2 representations are $K$-rational.

For $\lambda=2$ there are seven irreducible representations (see Table 4).
The representation $C_3$ is given by $C_3(S) = C_3(T) = - i$,
$C_4$ by $C_4(S) = C_4(T) = i$ and $R_2^0(1,3)_1$ is defined by
$ R_2^0(1,3) \cong R_2^0(1,3)_1\oplus C_1$.
All level 4 representations are $K$-rational.

For $\lambda=3$ there are 20 irreducible representations (see Table 5).
Here $\hat\chi$ is one of the two characters of $\Cal U$ of order 4 and
the representation $R_3^0(1,3,\chi_1)_1$ is defined by
$R_3^0(1,3,\chi_1) \cong R_3^0(1,3,\chi_1)_1
                         \oplus N_1(\chi_1) \oplus C_2\oplus C_2.$

For $\lambda=4$ there are  46 irreducible representations (see Table 6).
Here the representation $R_4^2(r,3,\chi_1)_1$ is given by the equality
$R_4^2(r,3,\chi_1) \cong R_4^2(r,3,\chi_1)_1 \oplus R_2^0(r,t)$.
\mn

\centerline{Table 3: Irreducible representations of $\Spl{2}{}$ }
\centerline{
\vbox{ \offinterlineskip
\def\Tablespace{ height2pt&\omit&&\omit&&\omit&&\omit&\cr }
\def\Tablerule{ \Tablespace
                \noalign{\hrule}
                \Tablespace      }
\hrule
\halign{&\vrule#&
  \strut\quad\hfil#\hfil\quad\cr
\Tablespace
& type of rep. &&  && dim  && \#  &\cr
\Tablerule
& $C_2 = N_1(\chi)$ &&  $\chi \in \P$     && $1$  && $1$      &\cr \Tablerule
& $N_1(\chi_1)$     && \omit              && $2$  && $1$      &\cr \Tablespace
}
\hrule}
}
\mn

\centerline{Table 4: Irreducible representations of $\Spl{2}{2}$}
\centerline{
\vbox{ \offinterlineskip
\def\Tablespace{ height2pt&\omit&&\omit&&\omit&&\omit&\cr }
\def\Tablerule{ \Tablespace
                \noalign{\hrule}
                \Tablespace      }
\hrule
\halign{&\vrule#&
  \strut\quad\hfil#\hfil\quad\cr
\Tablespace
& type of rep. &&  && dim && \# &\cr
\Tablerule
& $D_2(\chi)_+$ && $\chi \not\equiv 1$    && $3$  && $1$      &\cr \Tablerule
& $D_2(\chi)_-$ && $\chi \not\equiv 1$    && $3$  && $1$      &\cr \Tablerule
& $R_2^0(1,3)_1$ && \omit                 && $3$  && $1$      &\cr \Tablerule
& $C_2 \otimes R_2^0(1,3)_1$ &&\omit      && $3$  && $1$      &\cr \Tablerule
& $N_2(\chi)$ &&
  $\chi\in \P;\ \chi \not\equiv 1$         && $2$  && $1$      &\cr \Tablerule
& $C_3=R_2^0(3,1,\chi)$ &&
  $\chi\not\equiv 1$                      && $1$  && $1$      &\cr \Tablerule
& $C_4=R_2^0(1,1,\chi)$ &&
  $\chi\not\equiv 1$                      && $1$  && $1$      &\cr \Tablespace
}
\hrule}
}

\mn

\centerline{Table 5: Irreducible representations of $\Spl{2}{3}$}
\centerline{
\vbox{ \offinterlineskip
\def\Tablespace{ height2pt&\omit&&\omit&&\omit&&\omit&\cr }
\def\Tablerule{ \Tablespace
                \noalign{\hrule}
                \Tablespace      }
\hrule
\halign{&\vrule#&
  \strut\quad\hfil#\hfil\quad\cr
\Tablespace
& type of rep. &&  && dim  && \# &\cr
\Tablerule
& $D_3(\chi)_{\pm}$ &&
  $\chi\in \P$                       && $6$  && $4$      &\cr \Tablerule
& $R_3^0(1,3,\chi_1)_1$ &&\omit      && $6$  && $1$      &\cr \Tablerule
& $C_3 \otimes R_3^0(1,3,\chi_1)_1$ &&
  \omit                              && $6$  && $1$      &\cr \Tablerule
& $N_3(\chi)$ &&
  $\chi\in\P;\ \chi^2 \not\equiv 1$  && $4$  && $2$      &\cr \Tablerule
& $N_3(\chi)_{\pm}$ &&
  $\chi\in\P;\ \chi^2 \equiv 1$      && $2$  && $4$      &\cr \Tablerule
& $R_3^0(r,t,\hat\chi)$ &&
  $r=1,3;\ t=1,5$                    && $3$  && $4$      &\cr \Tablerule
& $R_3^0(1,t,\chi)_{\pm}$ &&
  $\chi\not\equiv 1;\ t=3,7$         && $3$  && $4$      &\cr \Tablespace
}
\hrule}
}

\mn

\centerline{Table 6: Irreducible representations of $\Spl{2}{4}$ }
\smallskip\noindent
\centerline{
\vbox{ \offinterlineskip
\def\Tablespace{ height2pt&\omit&&\omit&&\omit&&\omit&\cr }
\def\Tablerule{ \Tablespace
                \noalign{\hrule}
                \Tablespace      }
\hrule
\halign{&\vrule#&
  \strut\quad\hfil#\hfil\quad\cr
\Tablespace
& type of rep. &&  && dim  && \# &\cr
\Tablerule
& $D_4(\chi)$ && $\chi\in\P$                            && $24$  &&  $2$
&\cr \Tablerule
& $N_4(\chi)$ && $\chi\in\P$                            &&  $8$  &&  $6$
&\cr \Tablerule
& $R_4^0(r,t,\chi)$ &&
  $\chi\in\P;\ \chi\not\equiv 1;\ r=1,3;\ t=1,5$        &&  $6$  &&  $4$
&\cr \Tablerule
& $R_4^0(r,t,\chi)_\pm$ &&
  $\chi\in\P;\ \chi^2\equiv 1;\ r=1,3;\ t=1,5$            &&  $3$  && $16$
&\cr \Tablerule
& $R_4^0(1,t,\chi)_\pm$ && $\chi\in \P;\ t=3,7$         &&  $6$  &&  $8$
&\cr \Tablerule
& $R_4^2(r,t,\chi)$ &&
  $\chi\not\equiv 1;\ r,t \in \{1,3\}$                  &&  $6$  &&  $4$
&\cr \Tablerule
& $C_2 \otimes R_4^2(r,3,\chi)$ &&
  $\chi \not\equiv 1;\ r=1,3$                           &&  $6$  &&  $2$
&\cr \Tablerule
& $R_4^2(r,3,\chi_1)_1$ && $r=1,3$                      &&  $6$  &&  $2$
&\cr \Tablerule
& $N_3(\chi)_+\otimes R_4^0(1,7,\psi)_+$ &&
  $\chi\in\P;\ \chi^2\equiv1;\ \psi \not\equiv 1;$      && $12$  &&  $2$
&\cr \Tablespace
&\omit &&
   $\psi^2\equiv 1;\ \psi(-1) = 1$  && \omit && \omit
&\cr \Tablespace
}
\hrule}
}

\vfill\eject
For $\lambda=5$ there are 92 irreducible representations (see Table 7).
Here for fixed $r=1,3$ the 2 irreducible representations of type
$R_5^2(\cdot,1,\chi)_1$ ($ \chi\not\in\P$) are given by the 2
two dimensional irreducible level 5 subrepresentations of
$R^2_5(r,1)$.
\mn

\centerline{Table 7: Irreducible representations of $\Spl{2}{5}$ }
\smallskip\noindent
\centerline{
\vbox{ \offinterlineskip
\def\Tablespace{ height2pt&\omit&&\omit&&\omit&&\omit&\cr }
\def\Tablerule{ \Tablespace
                \noalign{\hrule}
                \Tablespace      }
\hrule
\halign{&\vrule#&
  \strut\quad\hfil#\hfil\quad\cr
\Tablespace
& type of rep. &&  && dim && \# &\cr
\Tablerule
& $D_5(\chi)$ && $\chi\in \P$            && $48$  &&  $4$      &\cr \Tablerule
& $N_5(\chi)$ && $\chi\in\P$             && $16$  && $12$      &\cr \Tablerule
& $R_5^0(r,t,\chi)$ &&
  $\chi\in\P;\ r=1,3;\ t=1,5$             && $12$  && $16$      &\cr \Tablerule
& $R_5^0(1,t,\chi)_\pm$ &&
  $\chi\in\P;\ t=3,7$                    && $24$  &&  $4$      &\cr \Tablerule
& $R_5^1(r,t,\chi)_\pm$ &&
  $\chi\in\P;\ r,t\in\{1,5\}\ \text{or}$ && $12$  && $16$      &\cr \Tablespace
& \omit && $r=1,3\ \text{and}\ t=3,7$ && \omit && \omit        &\cr \Tablerule
& $R_5^2(r,t,\chi)_\pm$ &&
  $\chi\in\P;\ r=1,3;\ t=1,3,5,7$        &&  $6$  && $32$      &\cr \Tablerule
& $R_5^2(r,1,\chi)_1$ &&
  $\chi\not\in\P;\ r=1,3$                && $12$  &&  $4$      &\cr \Tablerule
& $C_3 \otimes R_5^2(r,1,\chi)_1$ &&
  $\chi\not\in\P;\ r=1,3$                && $12$  &&  $4$      &\cr \Tablespace
}
\hrule}
}

\mn
For $\lambda>5$ there are the following irreducible representations (see
Table 8). Here $\chi$ are always primitive characters and
$R_\lambda^{\lambda-3}(r,t,\chi_{\pm 1})_1$ is the unique irreducible
level $2^\lambda$ subrepresentation of
$R_\lambda^{\lambda-3}(r,t,\chi_{\pm 1})$ which has dimension
$3\cdot 2^{\lambda-4}$.
\mn
\centerline{Table 8: Irreducible representations of $\Spl{2}{\lambda}$ for
                     $\lambda > 5$ }
\smallskip\noindent
\centerline{
\vbox{ \offinterlineskip
\def\Tablespace{ height2pt&\omit&&\omit&&\omit&&\omit&\cr }
\def\Tablerule{ \Tablespace
                \noalign{\hrule}
                \Tablespace      }
\hrule
\halign{&\vrule#&
  \strut\quad\hfil#\hfil\quad\cr
\Tablespace
& type of rep. \footnotemark &&  && dim  && \# &\cr
\Tablerule
& $D_\lambda(\chi)$ && \omit
  && $3 \cdot 2^{\lambda-1}$  &&  $2^{\lambda-3}$       &\cr \Tablerule
& $N_\lambda(\chi)$ && \omit
  && $2^{\lambda-1}$    && $3 \cdot 2^{\lambda-3}$      &\cr \Tablerule
& $R_\lambda^0(1,7,\chi)$ && $t=3,7$
  && $3 \cdot 2^{\lambda-2}$  && $2^{\lambda-3}$       &\cr \Tablerule
& $R_\lambda^\sigma(r,t,\chi)$ &&
  $\cases  r=1,3;\ t=1,5                     & \text{for}\ \sigma=0\\
           r,t\in \{1,5\}\ \text{or}\        & \\
           r=1,3\ \text{and}\ t=3,7         & \text{for}\ \sigma=1\\
           r=1,3;\ t=1,3,5,7                & \text{for}\ \sigma=2\\
   \endcases$
  && $3 \cdot 2^{\lambda-3}$  &&  $5 \cdot 2^{\lambda-2}$      &\cr \Tablerule
& $R_\lambda^\sigma(r,t,\chi)$ &&
  $\sigma=3,\dots,\lambda-3;\ r,t\in \{1,3,5,7\}$
  &&  $3 \cdot 2^{\lambda-4}$
  &&  $4\cdot \sum_{\sigma=3}^{\lambda-3} 2^{\lambda-\sigma}$
  &\cr \Tablerule
& $R_\lambda^{\lambda-2}(r,t,\chi)$ &&
  $r=1,3,5,7;\ t=1,3$
  && $3 \cdot 2^{\lambda-4}$  &&  $16$      &\cr \Tablerule
& $R_\lambda^{\lambda-3}(r,t,\chi_{\pm 1})_1$ &&
  $r=1,3,5,7;\ t= 1,3$
  && $3 \cdot 2^{\lambda-4}$  &&  $16$      &\cr \Tablespace
}
\hrule}
}
\footnotetext{For $\lambda=6$ one has to use representation of type
              $R_6^4(r,t,\chi_1)_1$ and
              $C_2\otimes R_6^4(r,t,\chi_1)_1$ ($r=1,3$)
              instead of those of type
              $R_\lambda^{\lambda-3}(r,t,\chi_{\pm 1})_1$.
              The representations $R_6^4(r,t,\chi_1)_1$ are the unique
              level 6 subrepresentations of  $R_6^4(r,t,\chi_1)$
              with dimension 12.
              }

\vfill\eject

\head
5. Results on the classification of strongly-modular fusion algebras
\endhead

\subhead
5.1 Classification of the strongly-modular fusion algebras of dimension less
    then or equal to four
\endsubhead

In this section we consider all two, three and four dimensional level $N$
representations of $\SL$ and investigate whether they
are admissible.

\proclaim{Main theorem 1}
Let $({\Cal F},\rho)$ be a two dimensional strongly-modular fusion algebra.
Then $({\Cal F},\rho)$ is isomorphic to the tensor product of
a one dimensional modular fusion algebra with one of the
modular fusion algebras in Table 9.
\endproclaim
\mn
\centerline{Table 9: Two dimensional strongly-modular fusion algebras }
\smallskip\noindent
\centerline{
\vbox{ \offinterlineskip
\def\Tablespace{ height2pt&\omit&&\omit&&\omit&\cr }
\def\Tablerule{ \Tablespace
                \noalign{\hrule}
                \Tablespace      }
\hrule
\halign{&\vrule#&
  \strut\quad\hfil#\hfil\quad\cr
\Tablespace
& $\Cal F$ && $\rho(S)$ && $\frac1{2\pi i }\log(\rho(T))\bmod \Z $  &\cr
\Tablerule
& $\Phi_1\cdot\Phi_1 = \Phi_0$
  &&  $\frac1{\sqrt2} \pmatrix -1 &-1 \\ -1 &1 \endpmatrix$
  &&  $\cases
        \diag( \frac18, \frac38) &\\
        \diag( \frac78, \frac58) &\\
       \endcases$
   & \cr  \Tablespace\Tablespace
& ( $\Z_2$ ) &&\omit &&\omit &\cr \Tablerule\Tablerule
& $\Phi_1\cdot\Phi_1 = \Phi_0 + \Phi_1$
       &&  $\frac{2}{\sqrt{5}}
            \pmatrix -\sin(\frac{\pi}{5}) & -\sin(\frac{2\pi}{5}) \\
                     -\sin(\frac{2\pi}{5}) &  \sin(\frac{\pi}{5})
            \endpmatrix$
       && $\cases
            \diag( \frac{19}{20}, \frac{11}{20} ) &\\
            \diag( \frac{1}{20}, \frac{9}{20} ) &\\
            \endcases$
  &\cr  \Tablespace \Tablespace\Tablespace
& ( "$(2,5)$" ) &&  $\frac{2}{\sqrt{5}}
            \pmatrix  -\sin(\frac{2\pi}{5}) & \sin(\frac{\pi}{5}) \\
                       \sin(\frac{\pi}{5})  & \sin(\frac{2\pi}{5})
            \endpmatrix$
       && $\cases
            \diag( \frac{3}{20}, \frac{7}{20} )&\\
            \diag( \frac{ 17}{20}, \frac{13}{20} )&\\
           \endcases$
  &\cr  \Tablespace}
\hrule}
}
\mn
\demo{Proof}
Let $({\Cal F},\rho)$ be a two dimensional strongly-modular fusion
algebra. Lemma 4 implies that $\rho$ is irreducible.
Therefore, we have to consider all irreducible two
dimensional representations of
$\SL$ which factor through a congruence subgroup.
By Lemma 7 we know that these representations can be obtained by
taking the tensor products of all irreducible two dimensional level
$p^\lambda$ representations with all one dimensional representations
of $\SL$.

There are exactly 11 inequivalent irreducible two dimensional
level $p^\lambda$ representations. Their explicit form is given in Appendix A.
We are interested in the classification of the two dimensional
strongly-modular fusion algebras up to tensor products with
one dimensional fusion algebras. Therefore, we can restrict our
investigation to one of the two dimensional representations of level
$2$, $2^3$, $3$ and the two representations of level 5 (see
Appendix A).
For the remaining 5 two dimensional representations
the eigenvalues of the image of $T$ are nondegenerate.
Hence, Lemma 2 implies that the corresponding matrix
representations are unique up to conjugation with
unitary diagonal matrices and permutation of the basis elements.
One can easily apply Verlinde's formula and check whether
the resulting coefficients $N_{i,j}^k$ have integer absolute
values for the two possible choices of
the basis element $\Phi_0$ corresponding to the vacuum
(conjugation with a unitary diagonal matrix does not change
the absolute value of $N_{i,j}^k$).
In particular for the level 2 representation $N_1(\chi_1)$ and the
level 3 representation $N_1(\chi)$ we obtain for both possible
choices of the distinguished basis elements $\Phi_0$ and $\Phi_1$
$$ \vert N_{1,1}^1 \vert =
   \cases \frac2{\sqrt{3}},& \qquad \text{for}\ N_1(\chi_1),\ p = 2 \\
          \frac1{\sqrt{2}},& \qquad \text{for}\ \ N_1(\chi),\ p = 3.
   \endcases$$
Since $\vert N_{1,1}^1\vert $ is not an integer we can
exclude these two representations.
For the level $2^3$ and $5$ representations one obtains integer
values for the $N_{i,j}^k$. Moreover, in all three cases both
possible choices of  the distinguished basis elements $\Phi_0$
and $\Phi_1$ lead to isomorphic fusion algebras.
We conclude that the representation of the modular group given by a
two dimensional strongly-modular fusion algebra is isomorphic to
the tensor product of a one dimensional representation  and
$N_3(\chi)_+$ ($p^\lambda = 2^3$) or $R_1(r,\chi_{-1})$
($r=1,2;p^\lambda = 5$).
Using that $\rho(S^2)$ should be a matrix consisting of
nonnegative integers one can determine the one dimensional
representation of $\SL$ up to an even one dimensional representation.
Therefore, $({\Cal F},\rho)$ is determined
up to tensor products with one dimensional modular fusion algebras.
The resulting representations and fusion algebras are collected
in Table 9.
\qed\enddemo

\remark{Remark}
The two fusion algebras in Table 9 are called $\Z_2$ and "$(2,5)$" fusion
algebras, respectively.
The first name is evident since this fusion algebra is isomorphic
to the group algebra of $\Z_2$ with the distinguished basis given
by the group elements. We will call the fusion algebra given by the
group algebra of $\Z_N$ in the following $\Z_N$ fusion algebra.
The second name results from the fact that
the Virasoro vertex operator algebra is rational
for $c = c(p,q) = 1-6\frac{(p-q)^2}{pq}$ ($p,q>1,\ (p,q)=1$)
\cite{\BPZ,\Wang} (these models are called Virasoro minimal models)
and the corresponding fusion algebra are denoted by "$(p,q)$" fusion algebra.
In particular the "$(2,5)$" fusion algebra is isomorphic to the
fusion algebra in the second row of Table 9.
\endremark
\mn
\proclaim{Main theorem 2}
Let $({\Cal F},\rho)$ be a three dimensional strongly-modular fusion algebra.
Then $({\Cal F},\rho)$ is isomorphic to the tensor product of
a one dimensional modular fusion algebra with one of the
modular fusion algebras in Table 10.
\endproclaim
\demo{Proof}
Let $({\Cal F},\rho)$ be a three dimensional strongly-modular
fusion algebra. By Lemma 7, $\rho$ is either irreducible
or isomorphic to a sum of a two dimensional and a one dimensional
irreducible representation.
We will now consider these two cases separately.
\sn
Firstly, assume that $\rho$ is irreducible. By Lemma 7, $\rho$ is
isomorphic to the tensor product of a one dimensional representation
and one of the three dimensional irreducible level $p^\lambda$
representations. There are exactly 33 inequivalent
irreducible 3 dimensional level $p^\lambda$ representations.
Their explicit form is given in Appendix A.
We are interested in the classification up to tensor products
with one dimensional modular fusion algebras. Therefore, we
can restrict our investigation to a set of irreducible representations which
are not related via
tensor products with one dimensional representations.
This means that we have to consider one representation of
level $3$ and $2^2$, two representations of level $5$ and $7$
and, finally, four representations of level $2^4$  (see Appendix A).

For these representations the eigenvalues of the image
of $T$ are nondegenerate so that we can proceed now as in the proof of
the Main theorem 1.

Using Verlinde's formula for the representation
$N_1(1,\chi_1)$ ($p=3$) we obtain $\vert N_{1,1}^1 \vert = \frac12$
for all possible choices of the distinguished basis.
In the same way one finds for $R_1(r,\chi_1)$ ($r=1,2;p=5$) that
$$ \cases \vert N_{1,1}^2 \vert = \frac1{\sqrt{2}}
          & \text{for}\
             \rho(T) = \diag(1,e^{2\pi i \frac{r}5},
                             e^{2\pi i \frac{4r}5}) \\
          & \ \text{or}\
             \rho(T) = \diag(1,e^{2\pi i \frac{4r}5},
                             e^{2\pi i \frac{r}5}) \\
         \vert N_{1,1}^1 \vert = \frac1{\sqrt{2}}
          & \text{for}\
            \rho(T) =  \diag(e^{2\pi i \frac{r}5},1,
                             e^{2\pi i \frac{4r}5}) \\
          \vert N_{1,1}^1 \vert = \frac1{\sqrt{2}}
          & \text{for}\
            \rho(T) =  \diag(e^{2\pi i \frac{4r}5},1,
                             e^{2\pi i \frac{r}5}). \\
\endcases $$
Here the different cases correspond to the different
possible choices of the distinguished basis.
We conclude that $\rho$ cannot be isomorphic to a tensor product of
a one dimensional representation and $N_1(1,\chi_1)$ ($p=3$)
or $R_1(r,\chi_1)$ ($r=1,2;p=5$).

\mn
\centerline{Table 10: Three dimensional strongly-modular fusion
                      algebras }
\smallskip\noindent
\centerline{
\vbox{ \offinterlineskip
\def\Tablespace{ height2pt&\omit&&\omit&&\omit&\cr }
\def\Tablerule{ \Tablespace
                \noalign{\hrule}
                \Tablespace      }
\hrule
\halign{&\vrule#&
  \strut\quad\hfil#\hfil\quad\cr
\Tablespace
& $\Cal F$ && $\rho(S)$ && $\frac1{2\pi i }\log(\rho(T)) \bmod \Z$  &\cr
\Tablerule
& $\Phi_1\cdot\Phi_1 = \Phi_2$
  &&  $$
  &&  $$
  &\cr \Tablespace\Tablespace\Tablespace
& $\Phi_1\cdot\Phi_2 = \Phi_0$
  &&  $\frac1{\sqrt{3}}\pmatrix 1 & 1                  & 1 \\
                                1 & e^{2\pi i \frac13} & e^{2\pi i \frac23} \\
                                1 & e^{2\pi i \frac23} & e^{2\pi i \frac13}
       \endpmatrix$
  &&  $ \diag( \frac14, \frac7{12}, \frac7{12})$
  &\cr \Tablespace\Tablespace\Tablespace
& $\Phi_2\cdot\Phi_2 = \Phi_1$
  &&  $$
  &&  $$
  &\cr \Tablespace\Tablespace
& ( $\Z_3$ )
  &&  $$
  &&  $$
  &\cr \Tablerule\Tablerule
& $\Phi_1\cdot\Phi_1 = \Phi_0 + \Phi_2$
 &&  $\frac2{\sqrt{7}} \pmatrix
      -s_2 &  -s_1 &   s_3 \\
       -s_1 & -s_3 &  -s_2 \\
       s_3 & -s_2 &    s_1
      \endpmatrix$
  &&  $\cases
        \diag( \frac47, \frac17, \frac27 ) &\\
        \diag( \frac37, \frac67, \frac57 ) &\\
       \endcases$
  &\cr \Tablespace\Tablespace\Tablespace
& $\Phi_1\cdot\Phi_2 = \Phi_1 + \Phi_2$
  &&  $\frac2{\sqrt{7}} \pmatrix
       -s_3 &   -s_1 &   s_2 \\
       -s_1  &  -s_2 &  -s_3 \\
        s_2 &  -s_3 &    s_1
      \endpmatrix$
  &&  $\cases
        \diag( \frac17, \frac47, \frac27)&\\
        \diag( \frac67, \frac37, \frac57)&\\
       \endcases$
  &\cr \Tablespace\Tablespace\Tablespace
& $\Phi_2\cdot\Phi_2 = \Phi_0 +\Phi_1 +\Phi_2$
 &&  $\frac2{\sqrt{7}} \pmatrix
       s_1 &  s_2 &  s_3 \\
      s_2 & -s_3 &  s_1 \\
      s_3 &   s_1 & -s_2
      \endpmatrix$
  &&  $\cases
        \diag( \frac27, \frac17, \frac47 )&\\
        \diag( \frac57, \frac67, \frac37 )&\\
       \endcases$
  &\cr \Tablespace\Tablespace\Tablespace
& ( "$(2,7)$" )   && $ s_j = \sin(\frac{j \pi}7)$
  &&  \omit
  &\cr \Tablerule\Tablerule
& $\Phi_1\cdot\Phi_1 = \Phi_0$
  &&  $$
  &&  $$
  &\cr \Tablespace\Tablespace\Tablespace
& $\Phi_1\cdot\Phi_2 = \Phi_2$
  &&  $\frac12\pmatrix 1         &1         & \sqrt{2}\\
                       1         &1         &-\sqrt{2}\\
                       \sqrt{2}  &-\sqrt{2} &0
       \endpmatrix$
  &&  $\cases
        \diag( \frac{8-n}{16}, \frac{16-n}{16}, \frac{n}8 ) &\\
        \diag( \frac{16-n}{16}, \frac{8-n}{16}, \frac{n}8 ) &\\
        n=0,\dots, 7 &\\
          \endcases$
  &\cr \Tablespace\Tablespace\Tablespace
& $\Phi_2\cdot\Phi_2 = \Phi_0 +\Phi_1$
  &&  $$
  &&  $$
  &\cr \Tablespace
& ( "$(3,4)$" )
  &&  $$
  &&  $$
  &\cr \Tablespace
}
\hrule}
}

\mn
An analogous calculation shows that for the representations of type
$R_1(r,\chi_{-1})$ one has $\vert N_{i,j}^k\vert \in \N$
for all 3 possible choices of the distinguished basis.
For the remaining representations one also
has $\vert N_{i,j}^k\vert \in \N$ for the two possible
choices of the distinguished basis (here the matrix $\rho(S)$
contains a zero so that there are only two possible choices
of the distinguished basis).

Hence, $\rho$ is isomorphic to a tensor product of a one
dimensional representation with one of these 7 representations.
Using that for a modular fusion algebra $\rho(S^2)_{i,j}$
equals $N_{i,j}^0$ one can determine the possible one dimensional
representations. The corresponding strongly-modular fusion
algebras are contained in Table 10 in the second and third row.

\mn
Secondly, assume that $\rho$ decomposes into a direct sum of
two irreducible representations $\rho \cong \rho_1 \oplus \rho_2$
with $\dim(\rho_j) = j$. Then $\rho_2$ is isomorphic to the
tensor product of a one dimensional representation with one of
the two dimensional irreducible level $p^\lambda$ representations
contained in Table A1.

Using Lemma 1 we conclude that $\rho(T)$ has degenerate
eigenvalues so that $\rho_2(T)$ must have an eigenvalue of the
form $e^{2\pi i \frac{n}{12}}$.
Hence, $\rho_2$ cannot be isomorphic to the tensor product of a one
dimensional representation and one of the two dimensional
irreducible level $5$ and $2^3$ representations in Table A1.
Using once more that $\rho(T)$ has degenerate eigenvalues
we obtain that $\rho$ is isomorphic to the tensor product of a
one dimensional representation with either $N_1(\chi_1)\oplus C_j$
($j=1,2;p=2$) or $N_1(\chi)\oplus B_j$ ($j=2,3;p=3$).
In order find out whether these four representations are
admissible we have to look for distinguished bases.

Let us first consider the case
$\rho \cong C\otimes(N_1(\chi)\oplus B_j)$
($j=2,3;p=3$) where $C$ is a one dimensional representation.
Here $\rho(S^2)$ has two different eigenvalues since $N_1(\chi)$
is odd and the representations $B_j$ are even.
Since the vacuum is selfconjugate, i.e.\ $\rho(S^2)_{0 0} = 1$
the representation $C$ has to be odd. Without loss of generality
we choose $C=C_4$ for $j=2$ and $C=C_3$ for $j=3$.
Furthermore, the fact that $\rho(S^2)$ has two different eigenvalues
implies that we must have
$$ \rho(S^2) = \pmatrix 1 &0 &0 \\ 0 &0 &1 \\ 0 &1 &0 \endpmatrix.$$
Using these two conditions it follows that in a basis in which $\rho(S^2)$
has this form and $\rho(T)$ is diagonal we must have
$$  \rho(S) = \frac1{\sqrt{3}}
        \pmatrix \epsilon & \epsilon           & \epsilon \\
                 \epsilon & e^{2\pi i \frac13} & e^{2\pi i \frac23} \\
                 \epsilon & e^{2\pi i \frac23} & e^{2\pi i \frac13}
       \endpmatrix,\qquad \epsilon^2 = 1
$$
and
$$\rho(T) = \cases \diag(e^{2\pi i\frac5{12}},e^{2\pi i\frac1{12}},
                         e^{2\pi i\frac1{12}}) &\text{or}  \\
                   \diag(e^{2\pi i\frac7{12}},e^{2\pi i\frac{11}{12}},
                        e^{2\pi i\frac{11}{12}}) &\\
            \endcases
$$
up to conjugation with a unitary diagonal matrix
(the two possibilities for $\rho(T)$ correspond  to
the two possible choices of the distinguished basis).

Applying now Verlinde's formula leads to a modular fusion algebra
iff $\epsilon=1$ for both choices of the distinguished basis.
The corresponding fusion algebra, $\rho(S)$ and
$\rho(T)$ are listed in the first row of Table 10.

Finally, consider the case
$\rho \cong C\otimes(N_1(\chi_1)\oplus C_j)$ ($j=1,2$).
Since $N_1(\chi_1)$ ($p=2$) and $C_j$ ($j=1,2$) are even $\rho$ has to be even,
too. Therefore, $C$ is even and w.l.o.g. we choose
$C=C_1$ for $j=1$ and $C=C_2$ for $j=2$.
Since $\rho$ is even one must have $\rho(S^2)= \id$ and,
therefore, $\rho(S)$ is real (c.f.\ the second remark in \S2).
Plugging this in we find (up to permutation of the basis elements)
that
$$\rho(S) = \frac12 \pmatrix 1          & -\sqrt{3}a & \sqrt{3}b \\
                             -\sqrt{3}a & 2-3a^2     & 3ab       \\
                              \sqrt{3}b & 3ab        & 3a^2-1
                    \endpmatrix, \qquad
  \rho(T) = (-1)^j \diag(1,-1,-1)
$$
where $a,b \in \R$ and $a^2+b^2 =1$.
Using Verlinde's formula we obtain as conditions for $\rho$ to
be admissible
$$ \cases  \frac{(1-3a^2)(3a^2-2)}{\sqrt{3}a} \in \N
           & \text{for}\ \rho(T) = (-1)^j \diag(1,-1,-1) \\
           \frac{1}{\sqrt{3} a (3a^2-2)},
           \frac{3a^2-1}{\sqrt{3} a (3a^2-2)} \in \N
           & \text{for}\ \rho(T) = (-1)^j \diag(-1,-1,1). \\
  \endcases
$$
The first case implies that $a^2 = \frac13$ or $a^2 = \frac23$
and the second one $a^2 = \frac13$, respectively. Inserting these values of
$a$ in the explicit form of $\rho(S)$ above we indeed obtain
modular fusion algebras if we choose the signs of $a$ and $b$
correctly. The resulting modular fusion algebras are contained
in the third row of Table 10. As fusion algebras they are
of type "$(3,4)$", also called Ising fusion algebra.

This completes the proof of the Main theorem 2.
\qed\enddemo

\proclaim{Main theorem 3}
Let $({\Cal F},\rho)$ be a four dimensional strongly-modular
fusion algebra. Then $({\Cal F},\rho)$ is either isomorphic to the
tensor product of 2 two dimensional strongly-modular fusion
algebras or isomorphic to the tensor product of a one
dimensional modular fusion algebra with one of the
modular fusion algebras in Table 11.
\endproclaim
\demo{Proof}
Let $({\Cal F},\rho)$ be a strongly-modular fusion algebra.
Then, by Lemma 4, we have the following possibilities for
$\rho$:
\roster
\item
$\rho$ is irreducible,
\item
$\rho\cong \rho_1\oplus\rho_2$ with $\dim(\rho_1) =3$, $\dim(\rho_2) = 1$,
\item
$\rho\cong \rho_1\oplus\rho_2$ with $\dim(\rho_1) = \dim(\rho_2) = 2$,
\item
$\rho\cong \rho_1\oplus\rho_2\oplus\rho_3$ with $\dim(\rho_1) =2$,
$\dim(\rho_2) = \dim(\rho_3) = 1$
\endroster
where $\rho_i$ ($i=1,2,3$) are irreducible representations.
\mn
\leftline{ $ \underline{ \text{(1) $\rho$ is irreducible}}$}
Assume that $\rho$ is irreducible. Then $\rho$
is either isomorphic to the tensor product of 2 two
dimensional representations of coprime levels  or it is
isomorphic to the tensor product of a one dimensional representation
with a four dimensional irreducible level $p^\lambda$ representation.
In the first case we obviously  have that $\rho$ is only admissible
iff both two dimensional representations are admissible (look at
Table A1). In this case the corresponding modular fusion algebra
is a tensor product of two fusion algebras contained in Table 9.
Let us now consider the other case, namely that
$\rho \cong C\otimes \rho_1$ where $C$ is a one dimensional
representation and $\rho_1$ is a four dimensional irreducible level
$p^\lambda$ representation.
In this case $\rho_1$ is given by one of the 9 representations
in Table A3. Note that for all of these representations the
eigenvalues of the image of $T$ are nondegenerate so that we
can use the argumentation used in the proof of the Main theorem 1.
\mn

\centerline{Table 11: Four dimensional simple strongly-modular fusion algebras}
\smallskip\noindent
\centerline{
\vbox{ \offinterlineskip
\def\Tablespace{ height2pt&\omit&&\omit&&\omit&\cr }
\def\Tablerule{ \Tablespace
                \noalign{\hrule}
                \Tablespace      }
\hrule
\halign{&\vrule#&
  \strut\quad\hfil#\hfil\quad\cr
\Tablespace
& $\Cal F$ && $\rho(S)$ && $\frac1{2\pi i }\log(\rho(T)) \bmod \Z$  &\cr
\Tablerule
& ${\displaystyle{ \Phi_1^2 = \Phi_2,\ \ \Phi_1\cdot\Phi_2 = \Phi_3,}
      \atop \displaystyle{} }
   \atop
   {\displaystyle{} \atop
    \displaystyle {
    \Phi_2^2 = \Phi_0,\ \  \Phi_1\cdot\Phi_3 = \Phi_0,} }$
 && $\frac12 \pmatrix 1&1&1&1\\ 1&i&-1&-i\\
                      1&-1&-1& -1\\ 1&-i& -1&i
             \endpmatrix$
 && $\cases
      \diag(\frac78, \frac14, \frac 38, \frac 14) &\\
      \diag(\frac38, \frac14, \frac 78, \frac 14) &\\
      \endcases$
 &\cr \Tablespace\Tablespace
& ${\displaystyle{ \Phi_3^2 = \Phi_2,\ \ \Phi_2\cdot\Phi_3 = \Phi_1,}
      \atop \displaystyle{} }
   \atop
   {\displaystyle{} \atop
    \displaystyle { (\ \Z_4 \ )} }$
 && $\frac12 \pmatrix 1&1&1&1\\ 1&-i&-1&i\\
                      1&-1&-1& -1\\ 1&i& -1&-i
             \endpmatrix$
 && $\cases
      \diag(\frac58, \frac34, \frac 18, \frac 34) &\\
      \diag(\frac18, \frac34, \frac 58, \frac 34) &\\
      \endcases$
 &\cr \Tablerule\Tablerule
& ${\displaystyle{ \Phi_1^2 = \Phi_0,\ \ \Phi_1\cdot\Phi_2 = \Phi_3,}
      \atop \displaystyle{\Phi_2^2 = \Phi_0,\ \  \Phi_1\cdot\Phi_3 = \Phi_2,} }
   \atop
   {\displaystyle{\Phi_3^2 = \Phi_0,\ \  \Phi_2\cdot\Phi_3 = \Phi_1 \,} \atop
    \displaystyle{} }$
 &&$\frac12 \pmatrix 1&-1&-1&-1 \\ -1&1&-1&-1 \\ -1&-1&1&-1 \\ -1&-1&-1&1
           \endpmatrix$
 &&$ \cases
   \diag(0,0,0,\frac12) &\\
   \diag(\frac12,0,0,0) &\\
   \endcases$
 &\cr \Tablespace\Tablespace
& ( $\Z_2 \otimes \Z_2$ ) && \omit && \omit
 &\cr \Tablerule\Tablerule
& ${\displaystyle{ \Phi_1^2 = \Phi_0+\Phi_3}  \atop \displaystyle{} }
   \atop
   {\displaystyle{} \atop
    \displaystyle {\Phi_1 \cdot \Phi_2 = \Phi_1+\Phi_3} }$
 && $ \frac23 \pmatrix
       -s_4 &  s_1 &  s_3 & -s_2 \\
        s_1 &  s_2 &  s_3 &  s_4 \\
        s_3 &  s_3 &    0 & -s_3 \\
       -s_2 &  s_4 & -s_3 &  s_1 \\
       \endpmatrix$
  && $\cases  \diag(\frac7{36},\frac{19}{36},\frac1{12},\frac{31}{36})  &\\
              \diag(\frac{29}{36},\frac{17}{36},\frac{11}{12},\frac5{36})  &\\
      \endcases $
 &\cr \Tablespace\Tablespace
& ${\displaystyle{ \Phi_1\cdot\Phi_3 = \Phi_2+\Phi_3}  \atop
    \displaystyle{} }
   \atop
   {\displaystyle{} \atop
    \displaystyle {  \Phi_2^2 = \Phi_0+\Phi_2+\Phi_3 } }$
 && $ \frac23 \pmatrix
       s_1 & s_2 & s_3 & s_4 \\
       s_2 &-s_4 & s_3 & -s1 \\
       s_3 & s_3 &   0 &-s_3 \\
       s_4 &-s_1 & -s_3 & s_2 \\
       \endpmatrix$
  && $\cases  \diag(\frac{31}{36},\frac7{36},\frac{1}{12},\frac{19}{36})  &\\
              \diag(\frac5{36},\frac{29}{36},\frac{11}{12},\frac{17}{36})  &\\
      \endcases $
 &\cr \Tablespace\Tablespace
& ${\displaystyle{ \Phi_2\cdot\Phi_3 = \Phi_1+\Phi_2+\Phi_3 }  \atop
    \displaystyle{} }
   \atop
   {\displaystyle{} \atop
    \displaystyle {  \Phi_3^2 = \Phi_0+\Phi_1+\Phi_2+\Phi_3 } }$
 && $ \frac23 \pmatrix
       s_2 &-s_4 & s_3 &-s_1 \\
      -s_4 & s_1 & s_3 &-s_2 \\
       s_3 & s_3 &   0 &-s_3 \\
      -s_1 &-s_2 &-s_3 &-s4 \\
       \endpmatrix$
  && $\cases \diag(\frac{19}{36},\frac{31}{36},\frac1{12},\frac7{36})  &\\
             \diag(\frac{17}{36},\frac{5}{36},\frac{11}{12},\frac{29}{12})  &\\
      \endcases $
 &\cr \Tablespace\Tablespace
& ( "$(2,9)$" )  && $ s_j = \sin(\frac{j\pi}9) $ && \omit
 &\cr \Tablespace
}
\hrule}
}

\mn
For the representation $N_1(\chi)$ ($\chi^3\not\equiv 1;p=5$) we
find by Verlinde's formula
$$ \vert N_{1,1 }^1 \vert  = \sqrt{3},\qquad \text{for}\ \
   \rho(T) = \diag( e^{2\pi i \frac{n}5},
                    e^{2\pi i \frac{3n}5},
                    e^{2\pi i \frac{2n}5},
                    e^{2\pi i \frac{4n}5})
   \qquad (n=1,\dots,4)
$$
where again the different possibilities for $\rho(T)$ correspond
to the different possible distinguished basis.
This shows that $\rho_1$ cannot be isomorphic to this representation.

Since the representation $N_1(\chi)$ ($\chi^3\equiv 1;p=5$) is
isomorphic to the tensor product of the two different level 5
representations in Table A1 it is clear that this representation is
admissible. Since the image of $T$ under this representation has nondegenerate
eigenvalues the corresponding modular fusion algebras
are isomorphic to the tensor product of 2 two dimensional modular
fusion algebras (as fusion algebras they are of type "$(2,5)$").

Consider now the representations $R_1(r,\chi_1)$ ($r=1,2;p=7$).
Here Verlinde's formula  implies that
$$ \vert N_{1,1}^1 \vert = \frac1{\sqrt{2}}\quad\text{for}\quad
   \rho(T) = \diag(e^{2\pi i \frac{n}7},1,\cdot,\cdot)\quad (n=1,\dots 6) $$
and
$$ \vert N_{1,1}^2 \vert  = \frac1{\sqrt{2}}\quad\text{for}\quad
   \rho(T) = \cases
     \diag(1,e^{2\pi i \frac27},e^{2\pi i \frac47},e^{2\pi i \frac17}) &
        \text{or} \\
     \diag(1,e^{2\pi i \frac57},e^{2\pi i \frac37},e^{2\pi i \frac67}).
    \endcases
$$
As above this removes these representations from the list of candidates
leading to modular fusion algebras.

For the representation  $N_3(\chi)$ ($\chi^3\not\equiv1; p=2^3$) one has
$$ \vert N_{1,1}^1 \vert = \sqrt{\frac43} \quad \text{for}\quad
   \rho(T) = \diag(e^{2\pi i \frac{2n+1}8},e^{2\pi i \frac{2n+5}8},\cdot,\cdot)
   \quad (n=1,\dots,4)
$$
so that this representation is also excluded.

Consider now the representations $R^1_2(r,1,\chi)$
($r=1,2; \chi^3\not\equiv 1; p=3^2$).
Here one has
$$ \vert N_{1,1}^1 \vert = \frac1{\sqrt3} \quad\text{for}\quad
   \rho(T) = \diag(e^{2\pi i \frac{r n^2}9},e^{2\pi i\frac{r}3},\cdot,\cdot)
   \quad (n=1,2,3).
$$
The basis element in the representation space corresponding to the
$\rho(T)$ eigenvalue of order three cannot correspond to $\Phi_0$
since in the corresponding row of $\rho(S)$ contains a zero.

Finally, the only remaining four dimensional irreducible level $p^\lambda$
representations that might lead to modular fusion algebras are those of type
$R_2^1(r,1,\chi)$ ($r=1,2; \chi^3\equiv 1;p^\lambda=3^2$).
Indeed, these representations lead to modular fusion algebras.
To be more precise one has to consider the tensor product of an odd
one dimensional representation with them because the $R_2^1(r,1,\chi)$
($\chi^3\equiv 1$) are odd themselves.
The corresponding fusion algebras are of type "$(2,9)$"
and the explicit form is given in Table 11.
The different modular fusion algebras result from the two
different representations and the fact that the distinguished
basis can be chosen in different ways.

\mn
\leftline{$\underline{\text{ $\rho\cong \rho_1\oplus\rho_2$ with
              $\dim(\rho_1) =3$, $\dim(\rho_2) = 1$}}$}
Assume that $\rho$ is isomorphic to the direct sum of a
one dimensional and an irreducible three dimensional representation.
Then one has $\rho \cong C\otimes ( \rho_1 \oplus D )$ where
$C$ and $D$ are one dimensional representations and $\rho_1$ is one
of the three dimensional irreducible level $p^\lambda$ representations
in Table A2. By Lemma 1 we know that $\rho(T)$ has degenerate
eigenvalues. Therefore, $\rho_1$ is of type $N_1(\chi_1)$ ($p=3$),
$R_1(r,\chi_1)$ ($r=1,2;p=5$), $D_2(\chi)_+$ ($p^\lambda=2^2$) or
$R_3^0(1,3)_{\pm}$ ($p^\lambda=2^3$).

Consider first the representation $N_1(\chi_1)$ ($p=3$).
In this case we can have $D=B_j$ ($j=1,2,3$).
Since $B_j$ and  $N_1(\chi_1)$ are even we can choose without loss of
generality $C=C_1$.
Using Verlinde's formula we find that
$$ \vert N_{1,1}^1 \vert = \frac12 \quad\text{for}\quad
   \rho(T) = \diag( e^{2\pi i \frac{j+1}3}, e^{2\pi i \frac{j+2}3},
                    e^{2\pi i \frac{j}3}, e^{2\pi i \frac{j}3})
$$
giving a contradiction for these choices of the distinguished basis.
For $\rho(T) = \diag( e^{2\pi i \frac{j}3},
                      e^{2\pi i \frac{j}3},
                      e^{2\pi i \frac{j+1}3},
                      e^{2\pi i \frac{j+2}3})$
the line of reasoning is a little bit more involved.
Here $N_{i,j}^0 = \rho(S^2)_{i,j} = \delta_{i,j}$ implies that
$\rho(S)$ is given by
$$ \rho(S) = \frac13\pmatrix 4b^2-1 &    4ab &  2a  &  2a \\
                                4ab & 3-4b^2 & -2b  & -2b \\
                                 2a &    -2b &  -1  &   2 \\
                                 2a &    -2b &   2  &   -1
                    \endpmatrix
$$
up to conjugation with an orthogonal diagonal matrix,
with $a,b \in \R$ and $a^2+b^2=1$.
With the explicit form of $\rho(S)$ we find as conditions for
$\rho$ to be admissible
$$  N_{1,1}^1 = \frac1{2a(3-4a^2)} \in \Z,\qquad
    N_{1,1}^2 = \frac{2a^2-1}{2a(3-4a^2)} \in \Z.
$$
However, the only solutions  that satisfy these two conditions
are those $a$ which equal $\frac1{2m}$ for an integer $m$ and
satisfy  $m^3 \equiv 0 \bmod 3m^2-1 $. It follows that
$ m \equiv 0 \bmod 3m^2-1 $ which gives a contradiction.
Therefore, the representations $N_1(\chi_1)\oplus B_j$ ($p=3$)
do not lead to modular fusion algebras.

Next we consider the representations $R_1(r,\chi_1)$ ($r=1,2;p=5$).
In this case the one dimensional representation $D$ has to be the
trivial one. Since these two representations are even we can
choose without loss of generality $C=C_1$, too. Using that $N_{i,j}^0 =
\delta_{i,j}$
we find that the matrix which describes the basis in the two
dimensional eigenspace corresponding to the eigenvalue
1 of $\rho(T)$ is orthogonal. Furthermore, by looking at suitable
$N_{i,j}^k$ we find that there are only two possibilities for
this matrix. In the corresponding basis we indeed find modular
fusion algebra given by the tensor product of two modular fusion
algebras of type "$(2,5)$".
That $\rho$ is admissible can also be interfered from the equality
$R_1(r,\chi_1)\oplus C \cong R_1(r,\chi_{-1}) \otimes
 R_1(r,\chi_{-1})$ ($r=1,2;p=5$).


Finally, we have to consider $D_2(\chi)_+$ ($p^\lambda=2^2$)
and $R_3^0(1,3,\chi)_{\pm}$ ($p^\lambda = 2^3$).
The corresponding possibilities for $\rho$ are
$ C_3 \otimes D_2(\chi)_+ \oplus C_j $ ($j=1,3,4$),
$ C_4 \otimes R_3^0(1,3,\chi)_+ \oplus C_3 $ or
$ C_3 \otimes R_3^0(1,3,\chi)_- \oplus C_4 $.
For the case  $\rho \cong  C_3 \otimes D_2(\chi)_+ \oplus C_1$
we obtain a modular fusion algebra given by the tensor product
of two $\Z_2$ fusion algebras. This can also be seen by looking
at the identity
$$C_3 \otimes D_2(\chi)_+ \oplus C_1 \cong
  D_2(\chi)_+ \otimes D_2(\chi)_+.$$
For $ C_4 \otimes R_3^0(1,3,\chi)_+ \oplus C_3 $ or
$ C_3 \otimes R_3^0(1,3,\chi)_- \oplus C_4 $ we obtain
$\Z_4$ type fusion algebras (see Table 11).
The other two representations
($ C_3 \otimes D_2(\chi)_+ \oplus C_j $ ($j=3,4$))
are not admissible as one can easily check by applying
Verlinde's formula.

\mn
\leftline{$\underline{\text{$\rho\cong \rho_1\oplus\rho_2$ with
              $\dim(\rho_1) = \dim(\rho_2) = 2$}}$}
Assume that $\rho$ decomposes into a direct sum of 2
two dimensional irreducible representations.
In this case we have
$\rho = C \otimes (\rho_1 \oplus D \otimes \rho_2)$ where
$C$ and $D$ are  one dimensional representations and $\rho_1,\rho_2$
are some level $p^\lambda$ representations contained in Table A1.
Since $\rho$ is reducible we know that $\rho(T)$ has
degenerate eigenvalues. This together with the parity of
the representations in Table A1 implies that $\rho$ equals
(up to a tensor product with an even one dimensional representation)
one of the following representations:
$$ \align
  & N_1(\chi_1) \oplus N_1(\chi_1)  \\
  &C_3 \otimes (N_1(\chi)   \oplus B_i\otimes N_1(\chi)) \quad
    (i=1,2) \\
  &C_4 \otimes (R_1(r,\chi_{-1})\oplus R_1(r,\chi_{-1})) \quad
    (r=1,2)\\
  &C_4 \otimes (N_3(\chi)_+ \oplus N_3(\chi)_+). \\
\endalign $$
In all cases we have that $\rho(S)$ is conjugate to a matrix
of block diagonal form. More precisely, this matrix consists
of two identical two by two matrices. A simple calculation
shows now that conjugation of $\rho(S)$ with a matrix which
leaves $\rho(T)$ diagonal leads to a matrix which has at least one
zero element in every row.
This is a contradiction since we have assumed that
$\rho$ is admissible and one can apply Verlinde's formula.

\mn
\leftline{$\underline{\text{ $\rho\cong \rho_1\oplus\rho_2\oplus\rho_3$
              with $\dim(\rho_1) =2$,
              $\dim(\rho_2) = \dim(\rho_3) = 1$}}$}
Assume that $\rho$ decomposes into a direct sum of an
irreducible two dimensional and 2 one dimensional representations.
Then, again by Lemma 1, $\rho(T)$ has degenerate eigenvalues
and a simple parity argument shows that the only possibilities
for $\rho$ are (up to a tensor product with an even one
dimensional representation):
$$ N_1(\chi_1) \oplus C_1 \oplus C_1 \qquad \text{or} \qquad
   N_1(\chi_1) \oplus C_1 \oplus C_2 $$
where $N_1(\chi_1)$ is the level 2 representation in Table A1.
We have to consider these two cases separately.

Firstly, let $\rho$ be conjugate to  $N_1(\chi_1) \oplus C_1 \oplus C_1$.
Then the requirements that $\rho(S)$ has to be symmetric and real
and that $\rho(T)$ has to be diagonal imply that
(up to permutation of the basis elements and conjugation with an
orthogonal diagonal matrix):
$$ \rho(S) = -\frac12
   \pmatrix
             -1&  \sqrt{3} a &  \sqrt{3} b &  \sqrt{3} c \\
    \sqrt{3} a & 3 a^2 - 2   &  3 a b      &  3 a c      \\
    \sqrt{3} b &  3 a b      &  3 b^2 - 2  &  3 b c      \\
    \sqrt{3} c &  3 a c      &  3 b c      &  3 c^2 - 2  \\
   \endpmatrix
$$
where $a,b,c\in \R$ with $a^2+b^2+c^2=1$ and
$\rho(T) = \diag(-1,1,1,1)$.

Fixing the distinguished basis such that $\Phi_0$ corresponds to the
eigenvector of $\rho(T)$ with eigenvalue $-1$  we obtain
$$ \align
   N_{1 1}^1 &= \frac{(2-3a^2)(1-3a^2)}{\sqrt{3}a},\ \
   N_{2 2}^2  = \frac{(2-3b^2)(1-3b^2)}{\sqrt{3}b},\ \
   N_{3 3}^3  = \frac{(2-3c^2)(1-3c^2)}{\sqrt{3}c} \\
   N_{1 1}^2 &= \sqrt{3} (3a^2-1)b,\qquad
   N_{1 1}^3  = \sqrt{3} (3a^2-1)c \\
   N_{2 2}^1 &= \sqrt{3} (3b^2-1)a,\qquad
   N_{2 2}^3  = \sqrt{3} (3b^2-1)c.
\endalign $$
This implies that $a^2 = b^2 = c^2 = \frac13$. The resulting
structure constants indeed define a fusion algebra, namely
the tensor product of two fusion algebras of type $\Z_2$.
As a modular fusion algebra this fusion algebra is
{\bf simple}, i.e.\ it is not a tensor product of two
nontrivial modular fusion algebras. The resulting modular
fusion algebra is contained in Table 11.

For the other choice of the distinguished basis where
$\Phi_0$ corresponds to an eigenvector $\rho(T)$ with
eigenvalue 1 we find
$$\align
 N_{3 3}^1 &= \frac{(3a^2-1)b}{a(3a^2-2)},\qquad
 N_{3 3}^2 = \frac{(3a^2-1)c}{a(3a^2-2)}, \\
 N_{3 3}^3 &= \frac{3a^2-1}{\sqrt{3}a(3a^2-2)},\qquad
 N_{2 2}^3 = \frac{1-3b^2}{\sqrt{3}a(3a^2-2)}
\endalign
$$
where the basis was chosen such that $\rho(T) = \diag(1,1,1,-1)$.
Let now $n := (N_{3 3}^1)^2 + (N_{3 3}^2)^2$ and
$m :=  (N_{3 3}^3)^2$.
It is now easy to verify that $n$ and $m$ satisfy the equation
$$ m^3 + (1-5 n) m^2 +(4 n^2+7 n) m + 4 n^2 - 3 n^3 = 0.$$
By Lemma 10 in \S 5.2 below the only nonnegative integer
solution of this equation is given by $n=m=0$.
Therefore, we find as the only possible solution $a^2=b^2=c^2=\frac13$.
The resulting structure constants define a fusion algebra
isomorphic to the tensor product of two $\Z_2$ fusion algebras.
However, analogous to the case of the other distinguished basis
discussed above this modular fusion algebra is {\bf simple}
and contained in Table 11.

Secondly, assume that $\rho$ is conjugate to
$N_1(\chi_1) \oplus C_1 \oplus C_2$.
Requiring that $\rho(S)$ is a symmetric real matrix and
that $\rho(T)$ is diagonal implies
(up to a permutation of the basis elements and conjugation
with an orthogonal diagonal matrix)
$$ \rho(S) = \frac12
\pmatrix
       3b^2-1 &        -3 a b & -\sqrt{3}  a c & \sqrt{3} a d \\
       -3 a b &        3a^2-1 & -\sqrt{3} b c  & \sqrt{3} b d \\
-\sqrt{3} a c & -\sqrt{3} b c &        3c^2-2  & -3 c d \\
 \sqrt{3} a d &  \sqrt{3} b d & -3 c d         & 3d^2-2\\
\endpmatrix
$$
where $a,b,c,d \in \R$ and $a^2+b^2 =1, c^2 + d^2 = 1$ and
$\rho(T) = \diag(1,1,-1,-1)$.
Using Verlinde's formula we obtain for the choice of the distinguished
basis in which $\Phi_0$ corresponds to the eigenvector of
$\rho(T)$ with eigenvalue 1
$$ ( N_{1 1}^1 )^2 =  \frac{(3a-1)^2(6a-5)^2}{9a^2(1-a^2)(3a-2)^2},\ \
   ( N_{1 1}^2 )^2 =  \frac{c^2}{3a^2(3a^2-2)},\ \
   ( N_{1 1}^3 )^2 =  \frac{d^2}{3a^2(3a^2-2)}.
$$
For the other choice of the distinguished basis
($\Phi_0$ corresponding to eigenvalue $-1$) one finds the
same expressions with $a$ and $c$ exchanged.

Let $n := ( N_{1 1}^2 )^2 + ( N_{1 1}^3 )^2$ and let
$m := ( N_{1 1}^1 )^2$. It is easy to verify that the
following equation for $n$ and $m$ holds true
$$ \align
   &(1-3n) m^3 +(12-37n+31n^2) m^2 + (48-152n+155n^2-53n^3) m \\
   &\quad    + 64 -208n +249n^2 -130n^3 + 25n^4 = 0.
\endalign
$$
By Lemma 11 in \S 5.2 below the only nonnegative integer
solution of this equation is given by  $m=0,n=1$.
This is a contradiction to the explicit form of $n$ and $m$
in terms of $a$ above.  Hence the representation
$N_1(\chi_1)\oplus C_1 \oplus C_2$ is not admissible.

This proves the Main theorem 3.
\qed\enddemo

\subhead
5.2 Proof of two Lemmas on diophantic equations
\endsubhead

\proclaim{Lemma 10}
Let $n$ be a nonnegative integer, $m$ a square of an integer
and $n,m$ solutions of
$$ m^3 + (1-5n) m^2 +(4n+7n^2) m + 4n^2-3n^3 = 0$$
Then $n=m=0$.
\endproclaim

\demo{Proof}
The equation can be written in the form
$$ (3n-m)(m-n)^2 = (m+2n)^2.$$
If $n=m$ then $m=n=0$. Otherwise, set
$t = \frac{m+2n}{m-n}$ implying
$$ m = \frac{(t+2)t^2}{2t-5},\qquad n = \frac{(t-1)t^2}{2t-5}. $$
If $m$ and $n$ are integral then also $t$ has to be integral
(any prime factor of the denominator of $t$ would divide the
denominator of $m$ and $n$).
Then $N=2t-5$ divides $(t-1)t^2 = \frac18 (N+5)^2(N+3)$ so that
$N$ divides $3\cdot 5^2$. None of the resulting 12 possibilities
leads to a nonnegative integer solution of $n,m$ where
$m\not=n$ and $m$ is a square.\qed
\enddemo
\mn
\proclaim{Lemma 11}
Let $n$ be a nonnegative integer, $m$ a square of an integer
and $n,m$ solutions of
$$ \align
   &(1-3n) m^3 +(12-37n+31n^2) m^2 + (48-152n+155n^2-53n^3) m \\
   &\quad    + 64 -208n +249n^2 -130n^3 + 25n^4 = 0.
\endalign
$$
Then $m=0,n=1$.
\endproclaim
\demo{Proof}
Set $k = m-n+4$, then the equation is equivalent to
$$  k^3 + 2 k^2 n - 3 k^3 n + 125 n^2 - 92 k n^2 + 22 k^2 n^2 -
    11 n^3 = 0.$$
If $k=0$ then $n=0$ and $m=-4$ is not a square. Otherwise, the equation
is equivalent to
$$ (-3 t+22 t^2) k^2 + (1+2t-92 t^2-11 t^3) k + 125 t^2 = 0,
    \quad k\not=0$$
where $t=\frac{n}{k}$. This equation has discriminant
$  (1 + 18 t + t^2) (1 - 7 t + 11 t^2)^2 $ and this must be a
square. Setting
$\frac{p}{q} := (1- t - (1 + 18 t + t^2)^{1/2})/(10t) \in \Q$
(with coprime $p,q$ and $q>0$) we get
$$ t =  \frac{ q (p + q)}{ p (5p + q)}. $$
Hence, using the quadratic equation in $k$ we finally have
$$ m = \frac{ (2p+q)^2 (p-q)^2}{p^2 (2q-p)^2 (p+q)},\qquad
   n = \frac{q^3}{p^2 (2q-p)}.
$$
The parametrization of $n$ implies that $p=\pm1$ and, furthermore,
that $q^3 \equiv 0 \bmod (2q-p)$. Therefore, we have
$p^3\equiv 0 \bmod (2q-p)$ so that $2q-p=\pm1$.
{}From the resulting four possiblities only $p=q=1$
satisfies the desired properties and leads to $m=0, n=1$.
\qed\enddemo

\remark{Remark}
Note that the proof of Lemma 10 and Lemma 11 relies essentially on
the fact that the curves defined by the two above equations are rational.
I would like to thank D.\ Zagier for discussion on the two lemmas
\cite{\Don}.
\endremark

\mn
\subhead
5.3 Classification of the nondegenerate strongly-modular fusion algebras with
    dimension less than 24
\endsubhead

In this section we classify all strong modular fusion algebras
$({\Cal F},\rho)$ of dimension less than 24 for which $\rho(T)$ has
nondegenerate eigenvalues. The main tool used in the proof
is the classification of the irreducible representations of the groups
$\Spl{p}{\lambda}$ described in section 4.

\proclaim{ Main theorem 4}
Let $({\Cal F},\rho)$ be a simple nondegenerate strongly-modular fusion
algebra.
Furthermore, assume that the dimension of $\Cal F$ is less than 24.
Then $\rho$ is isomorphic to the tensor product of an even one dimensional
representation of $\SL$ with one of the representations in Table 12.
Moreover, $\Cal F$ is isomorphic to $\Q[x]/<P(x)>$ with
distinguished basis $p_j(x)$ ($j=0,\dots,n-1$).
Here $P$ and $p_j$ are the unique polynomials satisfying
$$  \align
    P(x) &= \det( {\Cal N}_1 - x ) \\
    p_0(x) &= 1, \qquad p_1(x) = x,\qquad
    p_j(x) = \sum_{k=0}^{n-1} ({\Cal N}_1)_{j,k} \ p_k(x).
\endalign
$$
where the $({\Cal N}_1)_{j,k} := N_{1,j}^k$ are the fusion matrices
given in Appendix B.
\endproclaim
\mn
\centerline{Table 12: Simple nondegenerate strongly-modular fusion of dimension
                      less than 24}
\centerline{ ($q$ is a prime satisfying $q<47$)}
\smallskip\noindent
\centerline{
\vbox{ \offinterlineskip
\def\Tablespace{ height2pt&\omit&&\omit&&\omit&\cr }
\def\Tablerule{ \Tablespace
                \noalign{\hrule}
                \Tablespace      }
\hrule
\halign{&\vrule#&
  \strut\quad\hfil#\hfil\quad\cr
\Tablespace
&  fusion && dim && $\rho$
   &\cr \Tablerule
& $\Z_2$
   && $2$
   && $C_4\otimes N_3(\chi)_{\pm},\ (p^\lambda = 2^3)$
   &\cr \Tablerule
& "$c(3,4)$"
   && $3$
   && $ C_4\otimes D_2(\chi)_+,\quad (p^\lambda = 2^2) $
  &\cr \Tablespace
&  \omit    && \omit
   && $C_4\otimes R^0_3(1,3,\chi)_{\pm}, \quad (p^\lambda = 2^3)$
   &\cr \Tablespace
&  Ising      && \omit
   && $C_4\otimes R^0_4(r,3,\chi)_{\pm}, \ (r=1,2; p^\lambda = 2^4)$
   &\cr \Tablerule
&    "$(2,q)$"
   && $\frac12 (q-1)$
   && $ C_4^{\frac{q+1}2}\otimes
        R_1(r,\chi_{-1}),\ ( \bigl({r\over p}\bigr) = \pm 1;
                                p^\lambda = q  )$
   &\cr \Tablerule
&  "$(2,9)$"
   &&  4
   &&  $C_4\otimes R^1_2(r,1,\chi),\quad
        (r= 1,2; \chi^3\equiv 1; p^\lambda=3^2)$
   &\cr \Tablerule
& $\text{B}_{9}$
   &&  6
   && $N_2(\chi),\quad (\chi^3 \equiv 1; p^\lambda = 3^2)$
   &\cr \Tablerule
& $\text{B}_{11}$
   &&  10
   &&  $N_1(\chi),\quad (\chi^3\equiv 1; p^\lambda =11)$
   &\cr \Tablerule
& $\text{G}_9$
   && 12
   && $C_4\otimes R_3^1(r,1,\chi),\quad (r=1,2; \chi^3\equiv 1;p^\lambda=3^3)$
   &\cr \Tablerule
& $\text{G}_{17}$
   &&  16
   &&  $N_1(\chi),\quad (\chi^3\equiv 1; p^\lambda =17)$
   &\cr \Tablerule
& $\text{E}_{23}$
   && 22
   &&  $N_1(\chi),\quad (\chi^3\equiv 1; p^\lambda =23)$
   &\cr \Tablespace
}
\hrule
}
}
\remark{Remark}
For all fusion algebras in Table 12 apart from  $\text{B}_9$
there indeed exist RCFTs where the associated fusion
algebras are isomorphic to the ones in Table 12:
The fusion algebra in the first row occurs in the so-called $\Z_2$ model,
the ones in row 2, 3 and 4 in the
corresponding Virasoro minimal models (see also the remark
at the end of the Main Theorem 1) and, finally,
the ones in row 6, 7, 8 and 9  occur as fusion
algebras of certain rational models, so-called minimal models
of Casimir $\w$-algebras, namely for
${\Cal WB}_2$ and $c=-\frac{444}{11}$,
${\Cal WG}_2$ and $c=-\frac{590}{9}$,
${\Cal WG}_2$ and $c=-\frac{1420}{17}$ and
${\Cal WE}_7$ and $c=-\frac{3164}{23}$ \cite{\Ehof}.
The fusion algebras of type $\text{B}_9$ seems to
be related to ${\Cal WB}_2$ and $c=-24$. However, in
this case the model is not rational.
\endremark

\demo{Proof}
Let $({\Cal F},\rho)$ be a simple nondegenerate strongly-modular fusion algebra
of dimension less than 24. Lemma 1 implies that $\rho$ is irreducible.
Furthermore, since $({\Cal F},\rho)$ is strongly-modular we have
to consider all irreducible representations of $\Spl{N}{}$ of dimension
less than 24. Since $({\Cal F},\rho)$ is simple and nondegenerate
simple Lemma 7 shows that we can restrict our investigation to
irreducible representations of $\Spl{p}{\lambda}$.
Once again, since $({\Cal F},\rho)$ is nondegenerate we can follow
the line of reasoning in the proof of the Main theorem 1.

Therefore, we can directly apply Verlinde's formula to any such matrix
representation $\hat \rho$ and look  whether the resulting coefficients
$N_{i,j}^k$ have integer absolute values for the different choices
of the basis element corresponding to $\Phi_0$.
If the resulting numbers $N_{i,j}^k$ do not have integer absolute
values we can conclude that there exists no nondegenerate
strongly-modular fusion algebra $({\Cal F},\rho)$ where $\rho$ is
conjugate to the tensor product of a one dimensional representation
of $\SL$ and $\hat \rho$.
We have investigated this for all irreducible representations
of $\Spl{p}{\lambda}$ of dimension less than 24 by constructing them
explicitly\footnotemark{}.
\footnotetext{Here we have used the computer algebra system
PARI-GP \cite{\GP}.}

The proof of the theorem will consist of three separate cases:
We consider representations of  $\Spl{p}{}$ and $\Spl{p}{\lambda}$
and $\Spl{2}{\lambda}$ separately.
\mn
Firstly, let $\rho$ be isomorphic to a tensor product of a one dimensional
representation and an irreducible representation $\hat \rho$ of $\Spl{p}{}$
($p\not=2$).
Note that this case was already discussed in \cite{\Ehof}.

For the representations of type $D_1(\chi)$ the matrix $\rho(T)$ has
degenerate eigenvalues so that we can leave out this type of representation.

For the representations of type $N_1(\chi)$ we find
modular fusion algebras only for $p=5,11,17$ and $23$ and $\chi^3 \equiv 1$.
For $p=5$ the modular fusion algebra is not simple but equals
the tensor product of two modular fusion algebras where the
corresponding fusion algebras are of type "$(2,5)$"
(cf.\ also the proof of the Main theorem 3).
The modular fusion algebras corresponding to $p=11,17,23$ are
contained in the last three rows of Table 12.
As was already mentioned in \cite{\Ehof} these four representations
are probably the only admissible ones of type $N_1(\chi)$.
However, we do not have a proof of this statement but numerical checks
show that there is no other admissible representation of this type for
$p<167$~\cite{\Ehof}.

The representations of type $R_1(r,\chi_1)$ and $N_1(\chi_1)$ do not lead
to modular fusion algebras \cite{\Ehof}.

For all $\hat \rho$ of type $R_1(r,\chi_{-1})$ we obtain modular fusion
algebras. Here $\rho \cong  (C_4)^{\frac{p+1}2} \otimes R_1(r,\chi_{-1})$
is admissible for all odd primes $p$.
The corresponding modular fusion algebras are of type "$(2,p)$".
They are contained in the third row of Table~12.
\mn
Secondly, let  $\rho$ be isomorphic to a tensor product of a one dimensional
representation and a irreducible representation $\hat \rho$ of
$\Spl{p}{\lambda}$ ($p\not=2,\ \lambda>1$).

For the representations of type $D_\lambda(\chi)$ the matrix
$\rho(T)$ has degenerate eigenvalues excluding these representations
from our investigation.

The only representations of type $N_\lambda(\chi)$ which have dimension less
than 24 are those corresponding to $(p=3; \lambda=2,3)$ and $(p=5;\lambda=2)$.
A calculation shows that exactly one of these
representations leads to a modular fusion algebra.
This is the representation with $(p=3;\lambda=2)$ and $\chi^3\equiv1$.
The corresponding strongly-modular fusion algebra is contained in Table~12.

Only those representations of type $R_\lambda^\sigma(r,t,\chi)$ and
$R_\lambda(r,\chi_{\pm1})_1$ with $(p=3;\lambda=2,3)$ or $(p=5;\lambda=2)$
have dimension less than 24.
The representations $R_2^1(r,1,\chi)$ $(p^\lambda=3^2; \chi^3\equiv 1)$
lead to nondegenerate modular fusion algebras (cf.\ the proof of the
Main theorem 3).
{}From the other representations only those with
$p^\lambda=3^3;r=1,2;\chi^3\equiv1$ lead to modular
fusion algebras (see Table 12).
\mn
Thirdly, consider the irreducible representations of $\Spl{2}{\lambda}$.
All irreducible representations of dimension
less than or equal to 4 have been considered in the Main theorems 1 to 3.
The corresponding admissible representations with
nondegenerate eigenvalues of $\rho(T)$ are contained in Table 12.

For $\lambda=1,2$  all irreducible representations have dimension less
then
or equal to~3.

For $\lambda=3$ we have to consider the representations of
type $R_3^0(1,3,\chi_1)_1$ and $D_3(\chi)_{\pm}$.
The former representation  does not lead to a modular fusion algebra but
the representations $D_3(\chi)_{\pm}$ lead to modular fusion algebras of type
$\Z_2\otimes \text{"}(3,4)\text{"}$. The corresponding  modular fusion
algebras are composite and therefore not contained in Table 12.

For $\lambda=4$ only the irreducible representations of type
$R_4^0(r,t,\chi)_{\pm}$, $R_4^2(r,3,\chi_1)_1$ and $R_4^2(r,t,\chi)$
lead to modular fusion algebras. The first one leads to a fusion algebra
of type "$(3,4)$" (see Main theorem 2). The other two representations lead
to composite modular fusion algebras. These fusion algebras are of type
$\Z_2\otimes \text{"}(3,4)\text{"}$ and are not contained in Table 12.

For $\lambda=5,6$ there are no irreducible representation of dimension less
than 24 leading to modular fusion algebras
(some of them correspond to ``fermionic fusion algebras''
of $N=1$-Super-Virasoro minimal models which we do not discuss here).
\qed
\enddemo

\head
6. Conclusions
\endhead

In this paper we have classified all strongly-modular fusion algebras of
dimension less than or equal to four and all nondegenerate strongly-modular
fusion algebras of dimension less than 24.
In order to obtain our results we have used the classification of the
irreducible representations of the groups $\Spl{p}{\lambda}$.
Not all modular fusion algebras in our classification show up in known RCFTs.
However, all corresponding fusion algebras
are realized in known RCFTs apart from the fusion algebra
of type $\text{B}_9$. This fusion algebra can formally be related to
the Casimir $\w$-algebra ${\Cal WB}_2$ at $c=-24$ and seems to be an analogue
of the fusion algebra formally associated to the Virasoro
algebra with central charge $c = c(3,9)$.

\mn
\centerline{Table 13: Central charges and conformal dimensions}
\centerline{$\qquad\qquad$ of simple strongly-modular fusion algebras}
\smallskip\noindent
\centerline{
\vbox{ \offinterlineskip
\def\Tablespace{ height2pt&\omit&&\omit&&\omit&\cr }
\def\Tablerule{ \Tablespace
                \noalign{\hrule}
                \Tablespace      }
\hrule
\halign{&\vrule#&
  \strut\quad\hfil#\hfil\quad\cr
\Tablespace
& $\Cal F$    && $c \ (\bmod 4)$  && $h_i \ (\bmod\Z)$    &\cr \Tablerule
& $\Z_2$      && $1$          && $0,\frac14$  &\cr \Tablespace\Tablespace
& \omit       && $3$          && $0,\frac34$  &\cr \Tablerule
& $\Z_3$      && $2$          && $0,\frac13,\frac13$  or $0,\frac23,\frac23$
  &\cr \Tablerule
& $\Z_4$      &&  $1$ && $0,\frac18,\frac12,\frac18$ or
                         $0,\frac58,\frac12,\frac58$
  &\cr \Tablespace\Tablespace
& \omit       && $3$ && $0,\frac38,\frac12,\frac38$  or
                        $0,\frac78,\frac12,\frac78$
  &\cr \Tablerule
& $\Z_2\otimes\Z_2$  && $0$ && $0,0,0,\frac12$  or
                               $0,\frac12,\frac12,\frac12$
  &\cr \Tablerule
& "$(2,5)$"   && $\frac65$    && $0,\frac35$  &\cr \Tablespace\Tablespace
& \omit       && $\frac{14}5$ && $0,\frac25$  &\cr \Tablespace\Tablespace
& \omit       && $\frac25$    && $0,\frac15$  &\cr \Tablespace\Tablespace
& \omit       && $\frac{18}5$ && $0,\frac45$  &\cr \Tablerule
& "$(2,7)$"   && $\frac{16}7$ && $0,\frac47,\frac57$
  &\cr \Tablespace\Tablespace
& \omit       && $\frac{12}7$ && $0,\frac37,\frac27$
  &\cr \Tablespace\Tablespace
& \omit       && $\frac47$    && $0,\frac37,\frac17$
   &\cr \Tablespace\Tablespace
& \omit       && $\frac{24}7$ && $0,\frac47,\frac67$
   &\cr \Tablespace\Tablespace
& \omit       && $\frac87$    && $0,\frac67,\frac27$
  &\cr \Tablespace\Tablespace
& \omit       && $\frac{20}7$ && $0,\frac17,\frac57$
 &\cr \Tablerule
& "$(2,9)$"   && $\frac{10}3$ && $0,\frac13,\frac23,\frac{2n}9$
  &\cr \Tablespace\Tablespace
& \omit       && $\frac23$    && $0,\frac13,\frac23,\frac{n}9$
  &\cr \Tablespace\Tablespace
& \omit       && \omit        && $n=1,4,7$
  &\cr \Tablerule
& "$(3,4)$"   && $\frac{3n}2$ && $0,\frac12,\frac{n}{16}$
  &\cr \Tablespace\Tablespace
& \omit       && $n=0,\dots,15$ && \omit
  &\cr \Tablespace
}
\hrule}
}
\mn
The fact that we do not know examples of RCFTs for all of the
modular fusion algebras in our classification can be understood as follows.
The classification of the strongly-modular fusion algebras implies
restrictions on the central charge and the
conformal dimensions of possibly underlying RCFTs.
In Table 13 we have collected the possible values of $c$ and the $h_i$
for the simple strongly-modular fusion algebras of dimension less than or equal
to four. Note, however, that these restrictions are not as strong as the ones
in \cite{\Kir} for the two dimensional case or in \cite{\Caselle} for the
two and three dimensional case.
A natural way to obtain stronger restrictions
than the ones presented in Table 13 is to look whether there exist
vector valued modular functions transforming under the corresponding
representation of the modular group which have the correct pole order
at $i\infty$. This can be done using the methods developed in
\cite{\MIAU} and indeed leads to much stronger restrictions on $c$
and the $h_i$ as we will discuss elsewhere.
Of course, we expect that for any RCFT the corresponding characters
are modular functions so that these stronger restrictions have to be valid
explaining that our classification contains modular fusion algebras
for which we do not know of any realization in RCFTs.

{}From our considerations it is clear that a complete classification of
all simple nondegenerate strongly-modular fusion algebras is a purely
number theoretical problem which can probably be solved.

Unfortunately, the methods used in this paper seem to be not
sufficient for obtaining a complete classification of
strongly-modular fusion algebras. For those strongly-modular fusion algebras
which are not nondegenerate the corresponding representations of the modular
group are in general reducible and therefore there is a lot freedom
for possible choices of the distinguished basis in the representation
space. In the main theorems we have shown how one can
deal with this freedom in the case of two, three and four dimensional fusion
algebras.  However, we do not know a general method to overcome this
problem for arbitrary dimensions so that new methods have to be developed.

Finally, we would like stress that the main assumption for obtaining
our classifications, namely that fusion algebras are induced by
representations of $\Spl{N}{}$, is valid for all known examples
of rational conformal field theories.
Nevertheless, the question whether all fusion algebras associated to
RCFTs are strongly-modular is not yet answered.

\head
Acknowledgments
\endhead
I would like to thank  N.-P.\ Skoruppa for many discussions and
for drawing my attention to ref. \cite{\Nobs}.
Furthermore, I am grateful to  A.\ Recknagel, D.\ Zagier and
the research group of W.\ Nahm for many useful discussions.

We have used for many calculations the computer algebras system
PARI-GP~\cite{\GP}.

\vfill\eject
\head
7. Appendix A: The irreducible level $p^\lambda$  representations
               of dimension less than or equal to four
\endhead

Using the results in \S 4 one obtains as a complete list of two dimensional
irreducible level $p^\lambda$ representations
$$ \align
 &p^\lambda=2^1, \qquad N_1(\chi_1) \\
 &p^\lambda=3^1,\qquad N_1(\chi_1) \otimes B_i \\
 &p^\lambda=5^1,\qquad R_1(1,\chi_{-1}),\ R_1(2,\chi_{-1}) \\
 &p^\lambda=2^2, \qquad N_1(\chi_1)\otimes C_3 \\
 &p^\lambda=2^3, \qquad N_3(\chi)_+\otimes C_j \\
 & \text{where}\ i=1,2,3; \ j=1,\dots,4.
\endalign $$
The explicit form of the representations which are not related by tensor
products with $B_i$ or $C_j$ is given in  Table A1.
\mn
\centerline{Table A1: Two dimensional irreducible level $p^\lambda$
                      representations}
\smallskip\noindent
\centerline{
\vbox{ \offinterlineskip
\def\Tablespace{ height2pt&\omit&&\omit&&\omit&&\omit&\cr }
\def\Tablerule{ \Tablespace
                \noalign{\hrule}
                \Tablespace      }
\hrule
\halign{&\vrule#&
  \strut\quad\hfil#\hfil\quad\cr
\Tablespace
& level && type of rep. && $\rho(S)$ && $\frac1{2\pi i}\log(\rho(T)) $  &\cr
\Tablerule
& 2 && $N_1(\chi_1)$
  &&  $\frac1{2} \pmatrix -1 &-\sqrt{3} \\ -\sqrt{3} &1 \endpmatrix$
  &&  $\diag( 0,\frac12)$
  &\cr \Tablerule
& $3$ && $N_1(\chi)$
  &&  $-\frac{i}{\sqrt{3}} \pmatrix 1 &\sqrt{2} \\ \sqrt{2} &-1 \endpmatrix$
  &&  $\diag(\frac13, \frac23)$
  &\cr \Tablerule
& $5$ && $R_1(1,\chi_{-1})$
  &&  $\frac{2i}{\sqrt{5}}
       \pmatrix  -\sin(\frac{\pi}5)  & \sin(\frac{2\pi}5)\\
                  \sin(\frac{2\pi}5) & \sin(\frac{\pi}5)
       \endpmatrix$
  &&  $\diag( \frac15, \frac45)$
  &\cr \Tablespace\Tablespace
& \omit && $R_1(2,\chi_{-1})$
  &&  $ \frac{2i}{\sqrt{5}}
      \pmatrix  -\sin(\frac{2\pi}5)  & -\sin(\frac{\pi}5) \\
                -\sin(\frac{\pi}5)   & \sin(\frac{2\pi}5)
      \endpmatrix$
  &&  $\diag( \frac25, \frac35)$
  &\cr\Tablerule
& $2^3$ && $N_3(\chi)_{+}$
  &&  $\frac{i}{\sqrt{2}} \pmatrix -1 &-1 \\ -1 &1 \endpmatrix$
  &&  $\diag( \frac38, \frac58)$
  &\cr \Tablespace
}
\hrule}
}
\mn\mn
Similarly, one obtains as a complete list of three dimensional
irreducible level $p^\lambda$ representations
$$ \align
 &p^\lambda=3^1,\qquad N_1(\chi_1) \\
 &p^\lambda=5^1,\qquad R_1(1,\chi_1),\ R_1(2,\chi_1) \\
 &p^\lambda=7^1,\qquad R_1(1,\chi_{-1}),\ R_1(2,\chi_{-1}) \\
 &p^\lambda=2^2, \qquad D_2(\chi)_+\otimes C_j \\
 &p^\lambda=2^3, \qquad R^0_3(1,3,\chi)_+ \otimes C_j,\
                        R^0_3(1,3,\chi)_- \otimes C_j\\
 &p^\lambda=2^4, \qquad R^0_4(1,1,\chi)_+ \otimes C_j,\
                        R^0_4(1,1,\chi)_- \otimes C_j,\\
 &\qquad\qquad\qquad    R^0_4(3,1,\chi)_+ \otimes C_j,\
                        R^0_4(3,1,\chi)_- \otimes C_j\\
 & \text{where}\ j=1,\dots,4.
\endalign $$
The explicit form of the representations which are not related by tensor
products with $C_j$ is given in Table A2.

\vfill\eject
\centerline{Table A2: Three dimensional irreducible level
 $p^\lambda$  representations}
\smallskip\noindent
\centerline{
\vbox{ \offinterlineskip
\def\Tablespace{ height2pt&\omit&&\omit&&\omit&&\omit&\cr }
\def\Tablerule{ \Tablespace
                \noalign{\hrule}
                \Tablespace      }
\hrule
\halign{&\vrule#&
  \strut\quad\hfil#\hfil\quad\cr
\Tablespace
& level && type of rep. && $\rho(S)$ && $\frac1{2\pi i}\log(\rho(T))$
 &\cr \Tablerule
& 3 && $N_1(1,\chi_1)$
  &&  $ \frac13\pmatrix  -1 &2 &2 \\ 2 &-1 &2 \\ 2 &2 &-1 \endpmatrix$
  &&  $\diag( \frac13, \frac23, 0)$
  &\cr \Tablerule\Tablerule
& 5 && $R_1(1,\chi_1)$
  &&  $ \frac2{\sqrt{5}}\pmatrix
              \frac12 &\frac1{\sqrt2}      &\frac1{\sqrt2} \\
       \frac1{\sqrt2} &-s_1  &s_2 \\
       \frac1{\sqrt2} & s_2 &-s_1
       \endpmatrix$
  &&  $\diag(0,\frac15, \frac45 )$
  &\cr \Tablespace\Tablespace
& \omit && $R_1(2,\chi_1)$
  &&  $ \frac2{\sqrt{5}}\pmatrix
              -\frac12 &-\frac1{\sqrt2}      &-\frac1{\sqrt2} \\
       -\frac1{\sqrt2} &-s_2   &s_1 \\
       -\frac1{\sqrt2} & s_1   &-s_2
       \endpmatrix$
  &&  $\diag( 0, \frac25, \frac35 )$
  &\cr \Tablespace
&\omit && \omit && $s_j = \cos(\frac{j\pi}5)$ && \omit
  &\cr \Tablerule\Tablerule
& 7 && $R_1(1,\chi_{-1})$
  &&  $\frac2{\sqrt{7}} \pmatrix
       s_1 &  s_2 &  s_3 \\
      s_2 & -s_3 &  s_1 \\
      s_3 &   s_1 & -s_2
      \endpmatrix$
  &&  $\diag( \frac27, \frac17, \frac47 )$
  &\cr \Tablespace\Tablespace
& \omit && $R_1(2,\chi_{-1})$
  &&  $ -- \text{"} -- $
  &&  $\diag( \frac57, \frac67, \frac37 )$
  &\cr \Tablespace
&\omit && \omit && $s_j = \sin(\frac{j\pi}7)$ && \omit
  &\cr \Tablerule\Tablerule
& $2^2$ && $D_2(\chi)_+$
  &&  $ \frac{i}2\pmatrix  0        &\sqrt{2} &\sqrt{2} \\
                           \sqrt{2} &-1        &1       \\
                           \sqrt{2} &1         &-1
       \endpmatrix$
  &&  $\diag(\frac14, \frac12, 0 )$
 &\cr \Tablerule\Tablerule
& $2^3$ && $R^0_3(1,3,\chi)_+$
  &&  $ \frac{i}2\pmatrix  0        &\sqrt{2} &\sqrt{2} \\
                           \sqrt{2} &1        &-1       \\
                           \sqrt{2} &-1       &1
       \endpmatrix$
  &&  $\diag( \frac12, \frac58, \frac18 )$
 &\cr\Tablespace\Tablespace
& \omit && $R^0_3(1,3,\chi)_-$
  &&   $ (-1)\ \cdot (--\text{"}--) $
  &&  $\diag( \frac12, \frac78,\frac38 )$
 &\cr\Tablerule\Tablerule
& $2^4$ && $R^0_4(1,1,\chi)_{+}$
  &&  $\frac{i}2\pmatrix  0        &\sqrt{2} &\sqrt{2} \\
                          \sqrt{2} &1        &-1       \\
                          \sqrt{2} &-1       &1
       \endpmatrix$
  &&  $\diag(\frac{5}{8}, \frac{1}{16}, \frac{9}{16} )$
&\cr\Tablespace\Tablespace
& \omit && $R^0_4(1,1,\chi)_{-}$
  &&  $-- \text{"} --$
  &&  $\diag( \frac{1}{8}, \frac{5}{16}, \frac{13}{8} )$
&\cr\Tablespace\Tablespace
& \omit && $R^0_4(3,1,\chi)_{+}$
  &&  $\frac{i}2\pmatrix  0        &\sqrt{2} &\sqrt{2} \\
                          \sqrt{2} &-1       &1        \\
                          \sqrt{2} &1        &-1
       \endpmatrix$
  &&  $\diag( \frac{7}{8}, \frac{3}{16}, \frac{11}{16} )$
&\cr\Tablespace\Tablespace
& \omit && $R^0_4(3,1,\chi)_{-}$
  &&  $ -- \text{"} -- $
  &&  $\diag( \frac{3}{8}, \frac{15}{16}, \frac{7}{16} )$
 &\cr\Tablespace
}
\hrule}
}
\vfill\eject
\centerline{Table A3: Four dimensional irreducible level
                      $p^\lambda$ representations}
\smallskip\noindent
\centerline{
\vbox{ \offinterlineskip
\def\Tablespace{ height2pt&\omit&&\omit&&\omit&&\omit&\cr }
\def\Tablerule{ \Tablespace
                \noalign{\hrule}
                \Tablespace      }
\hrule
\halign{&\vrule#&
  \strut\quad\hfil#\hfil\quad\cr
\Tablespace
& level && type of rep. && $\rho(S)$ && $\frac1{2\pi i}\log(\rho(T))$  &\cr
\Tablerule
& 5 && $N_1(\chi),\ \chi^3 \not\equiv 1$
  && $\frac{2i}5
       \pmatrix  \eta_-       & \sqrt{3} s_2 & \eta_+       & \sqrt{3} s_4 \\
                 \sqrt{3} s_2 &  -\eta_+     & \sqrt{3} s_4 & \eta_- \\
                 \eta_+       & \sqrt{3} s_4 & -\eta_-      & -\sqrt{3} s_2 \\
                 \sqrt{3} s_4 & \eta_-       & -\sqrt{3} s_2& \eta_+  \\
       \endpmatrix$
  &&  $\diag( \frac35,\frac45,\frac25,\frac15 )$
 &\cr \Tablespace\Tablespace
&\omit && \omit
 && $ s_j = \sin(\frac{j\pi}5),\
      \eta_{\pm} = s_2 \pm s_4$
 && \omit
  &\cr \Tablespace\Tablespace
& \omit && $N_1(\chi),\ \chi^3 \equiv 1$
  && $-\frac25
       \pmatrix  \xi_1 & -\xi_2 & \xi_1 & -\xi_3  \\
                -\xi_2 & -\xi_1 & \xi_3 & \xi_1 \\
                 \xi_1 &  \xi_3 & \xi_1 & \xi_2 \\
                -\xi_3 &  \xi_1 & \xi_2 & -\xi_1  \\
       \endpmatrix$
  &&  $\diag( \frac35, \frac45,\frac25,\frac15 )$
  &\cr \Tablespace\Tablespace
&\omit && \omit
 && $  r_j = \cos(\frac{j\pi}5),\
      \xi_1 = r_1-r_4-\frac12,$
 && \omit
  &\cr \Tablespace
&\omit && \omit
 && $ \xi_2 = 3r_2+ 2r_4,\
      \xi_3 = 2r_2+ 3r_4$
 && \omit
  &\cr \Tablerule\Tablerule
& 7 && $R_1(1,\chi_1) $
  && $\sqrt{\frac{2}7}i \pmatrix
  -\frac1{\sqrt{2}} & -1     &   -1  &   -1 \\
  -1               &  \xi_1 & \xi_2 & \xi_3 \\
  -1               &  \xi_2 & \xi_3 & \xi_1 \\
  -1               &  \xi_3 & \xi_1 & \xi_2 \\
       \endpmatrix$
  &&  $ \diag( 0,\frac17,\frac47,\frac27 )$
  &\cr \Tablespace\Tablespace
& \omit && $R_1(2,\chi_1) $
  && $(-1) \cdot\qquad (-- \text{"} -- )$
  &&  $ \diag(0,\frac67,\frac37,\frac57)$
  &\cr \Tablespace\Tablespace
&\omit && \omit
 && $  s_j = \sqrt{\frac27}\sin(\frac{j\pi}7), $
&& \omit
  &\cr \Tablespace
&\omit && \omit
 && $ \xi_1 = 2s_2-s_4,\
      \xi_2 = 2s_4+s_6 $
&& \omit
  &\cr \Tablespace
&\omit && \omit
 &&  $\xi_2 = 2s_4+s_6,\
      \xi_3 = -2s_6-s_2 $
 && \omit
  &\cr \Tablerule\Tablerule
& $2^3$ && $N_3(\chi),\ \chi^3 \not\equiv 1 $
  && $ \frac{i}{\sqrt{8}} \pmatrix
   1 &  1 &  \sqrt{3} i & -s_1 \sqrt{3}i \\
   1 & -1 & -\sqrt{3} i & -s_1 \sqrt{3}i  \\
   -\sqrt{3} i  & \sqrt{3} i & 1 &  s_1 \\
  s_2 \sqrt{3}i & s_2 \sqrt{3} i & s_2 & -1 \\
       \endpmatrix$
  &&  $\diag(\frac38,\frac58,\frac18,\frac78) $
  &\cr \Tablespace\Tablespace
&\omit && \omit
 && $ s_j = e^{2\pi i \frac{j}3} $
 && \omit
  &\cr \Tablerule\Tablerule
& $3^2$ && $R^1_2(1,1,\chi),\ \chi^3 \equiv 1 $
  && $ \frac{2i}3\pmatrix
       -s_8 & -s_4 & -s_2 & -s_6 \\
       -s_4 &  s_2 & -s_8 &  s_6 \\
       -s_2 & -s_8 &  s_4 &  s_6 \\
       -s_6 &  s_6 &  s_6 &    0 \\
       \endpmatrix$
  &&  $\diag(\frac49,\frac19,\frac79,\frac13) $
  &\cr \Tablespace\Tablespace
& \omit && $R^1_2(2,1,\chi),\ \chi^3 \equiv 1 $
  && $ (-1)\ \cdot \quad (-- \text{"} --) $
  &&  $\diag( \frac29,\frac59,\frac89,\frac23 )$
  &\cr \Tablespace\Tablespace
& \omit && $R^1_2(1,1,\chi),\ \chi^3 \not\equiv 1 $
  && $ \frac{2}3\pmatrix
        s_1 &  s_5 &  s_7 &  s_6 \\
        s_5 & -s_7 & -s_1 &  s_6 \\
        s_7 & -s_1 &  s_5 & -s_6 \\
        s_6 &  s_6 &  -s_6 &    0 \\
       \endpmatrix$
  &&  $\diag( \frac49,\frac19,\frac79,\frac13 )$
  &\cr \Tablespace\Tablespace
& \omit && $R^1_2(2,1,\chi),\ \chi^3 \not\equiv 1 $
  &&  $ -- \text{"} -- $
  &&  $\diag( \frac59,\frac89,\frac29,\frac23) $
  &\cr \Tablespace\Tablespace
&\omit && \omit
 && $ s_j = \sin(\frac{\pi j}{18}) $
 && \omit &\cr\Tablespace
}
\hrule}
}
\vfill
\eject

Similarly, one obtains as a complete list of four dimensional
irreducible level $p^\lambda$ representations
$$ \align
 &p^\lambda=5^1,\qquad N_1(\chi)\ (\chi^3\not\equiv 1),
                       N_1(\chi)\ (\chi^3\equiv 1),  \\
 &p^\lambda=7^1,\qquad R_1(1,\chi_1),\ R_1(2,\chi_1) \\
 &p^\lambda=2^3,\qquad N_3(\chi),\ C_4\otimes  N_3(\chi) \\
 &p^\lambda=3^2,\qquad B_i \otimes R^1_2(1,1,\chi),
                       B_i \otimes R^1_2(2,1,\chi)
\endalign $$
where $i=1,2,3$ and for $p^\lambda=3^2$ the character $\chi$ is a
primitive character of order 3 or 6 (so there are 12 four dimensional
irreducible level $3^2$ representations).

The explicit form of the representations which are not related by tensor
products with $C_j$ or $B_i$ is given in  Table A3.
\head
8. Appendix B: Fusion matrices and graphs of the nondegenerate
   strongly-modular fusion algebras of dimension less than 24
\endhead
The fusion matrices ${\Cal N}_1$ which define the distinguished basis
of the simple nondegenerate strongly-modular fusion algebras of
dimension less than 24 are given by:
$$\align
   &\Z_2:\qquad\quad
    {\Cal N}_1 = \pmatrix 0 & 1 \\ 1 & 0 \endpmatrix  \\
   &\text{"} (3,4) \text{"}: \quad
    {\Cal N}_1 = \pmatrix 0 & 0 & 1 \\ 0 & 0 & 1 \\ 1 & 1 & 0 \endpmatrix \\
   &\text{"} (2,q) \text{"}:\quad
    {\Cal N}_1 = \left. \pmatrix  0 & 1       &         &     \\
                           1 & \ddots  &  \ddots &     \\
                             & \ddots  &   0     &  1  \\
                             &         &   1     &  1  \\
                \endpmatrix  \right\} \scriptstyle{\frac{q-1}{2}}\\
  & \text{B}_{9}:\ \qquad
    {\Cal N}_1 =  \pmatrix 0&1&0& & & \\
                           1&0&1&1& & \\
                           0&1&0&0&1& \\
                            &1&0&1&1&0\\
                            & &1&1&1&1\\
                            & & &0&1&1
                  \endpmatrix \\
 &\text{B}_{11}:\qquad
    {\Cal N}_1  = \pmatrix 0&1&0&0& & & & & & \\
                           1&0&1&0&0& & & & & \\
                           0&1&0&0&1&0& & & & \\
                           0&1&0&0&1&0&1& & & \\
                            &0&1&1&0&1&0&1& & \\
                            & &0&0&1&0&0&0&1& \\
                            & & &1&0&0&0&1&0&0\\
                            & & & &1&0&1&1&1&0\\
                            & & & & &1&0&1&1&1\\
                            & & & & & &0&0&1&1\\
                 \endpmatrix \\
  \endalign $$
$$\align
&\text{G}_9:\qquad
  {\Cal N}_1 =
   \left(\smallmatrix
0& 1& 0& 0& 0&  &  &  &  &  &  &  \\
1& 1& 1& 1& 0& 0&  &  &  &  &  &  \\
0& 1& 1& 1& 1& 1& 0&  &  &  &  &  \\
0& 1& 1& 0& 0& 1& 0& 0&  &  &  &  \\
0& 0& 1& 0& 1& 1& 1& 1& 0&  &  &  \\
 & 0& 1& 1& 1& 1& 0& 1& 1& 0&  &  \\
 &  & 0& 0& 1& 0& 1& 1& 0& 1& 1&  \\
 &  &  & 0& 1& 1& 1& 1& 1& 0& 1& 1\\
 &  &  &  & 0& 1& 0& 1& 0& 0& 0& 1\\
 &  &  &  &  & 0& 1& 0& 0& 1& 1& 0\\
 &  &  &  &  &  & 1& 1& 0& 1& 1& 1\\
 &  &  &  &  &  &  & 1& 1& 0& 1& 1\\
   \endsmallmatrix\right) \\
&\text{G}_{17}:\qquad
  {\Cal N}_1 =
   \left(\smallmatrix
0&1&0&0&0&0&0&0&0& & & & & & & \\
1&1&0&1&1&0&0&0&0&0& & & & & & \\
0&0&0&0&0&1&0&0&0&1&0& & & & & \\
0&1&0&0&1&0&0&0&0&0&1&0& & & & \\
0&1&0&1&1&0&0&0&0&0&1&1&0& & & \\
0&0&1&0&0&0&0&0&0&1&0&0&0&1& & \\
0&0&0&0&0&0&0&0&1&1&0&0&0&0&1& \\
0&0&0&0&0&0&0&0&0&0&1&0&0&1&0&1\\
0&0&0&0&0&0&1&0&1&0&0&0&1&0&1&0\\
 &0&1&0&0&1&1&0&0&1&0&0&0&1&1&0\\
 & &0&1&1&0&0&1&0&0&1&1&0&0&0&1\\
 & & &0&1&0&0&0&0&0&1&1&1&0&0&1\\
 & & & &0&0&0&0&1&0&0&1&1&0&1&1\\
 & & & & &1&0&1&0&1&0&0&0&1&1&1\\
 & & & & & &1&0&1&1&0&0&1&1&1&1\\
 & & & & & & &1&0&0&1&1&1&1&1&1\\
\endsmallmatrix\right) \\
\endalign
$$
and finally for  $\text{E}_{23}$ the matrix ${\Cal N}_1$ is given by
$$
\left( \smallmatrix
0&1&0&0&0&0&0&0&0&0&0&0&0&0&0& & & & & & & \\
1&0&1&1&0&0&1&0&0&0&0&0&0&0&0&0& & & & & & \\
0&1&0&0&1&0&0&0&0&1&0&0&0&0&0&0&0& & & & & \\
0&1&0&1&0&0&0&0&0&1&0&0&0&0&0&1&0&0& & & & \\
0&0&1&0&0&0&1&1&0&0&0&0&0&0&0&0&0&0&1& & & \\
0&0&0&0&0&0&0&0&0&1&0&1&0&0&1&0&0&0&0&0& & \\
0&1&0&0&1&0&0&0&0&1&0&0&0&1&0&1&0&0&0&0&0& \\
0&0&0&0&1&0&0&0&0&1&0&0&0&1&1&0&0&0&0&0&0&1\\
0&0&0&0&0&0&0&0&1&0&0&0&1&0&1&0&0&0&0&0&1&0\\
0&0&1&1&0&1&1&1&0&0&0&0&0&0&0&1&0&0&1&1&0&0\\
0&0&0&0&0&0&0&0&0&0&0&1&1&0&1&0&0&1&0&0&0&1\\
0&0&0&0&0&1&0&0&0&0&1&0&0&0&0&1&0&0&0&1&1&0\\
0&0&0&0&0&0&0&0&1&0&1&0&1&0&0&0&1&1&0&0&1&0\\
0&0&0&0&0&0&1&1&0&0&0&0&0&0&0&0&1&0&1&1&0&1\\
0&0&0&0&0&1&0&1&1&0&1&0&0&0&0&0&1&0&1&1&1&0\\
 &0&0&1&0&0&1&0&0&1&0&1&0&0&0&1&0&0&1&1&0&1\\
 & &0&0&0&0&0&0&0&0&0&0&1&1&1&0&1&1&0&0&1&1\\
 & & &0&0&0&0&0&0&0&1&0&1&0&0&0&1&1&0&1&1&1\\
 & & & &1&0&0&0&0&1&0&0&0&1&1&1&0&0&1&0&1&1\\
 & & & & &0&0&0&0&1&0&1&0&1&1&1&0&1&0&1&1&1\\
 & & & & & &0&0&1&0&0&1&1&0&1&0&1&1&1&1&1&1\\
 & & & & & & &1&0&0&1&0&0&1&0&1&1&1&1&1&1&1\\
\endsmallmatrix
\right).
$$
The fusion graphs corresponding to the fusion matrices ${\Cal N}_1$
can be found in a seperate postscript file.

\vfill\eject


\Refs
\refstyle{A}
\widestnumber\key{xx}

\ref\key \Schellekens
\by A.\ N.\  Schellekens
\paper Meromorphic $c=24$ Conformal Field Theories
\jour Commun. Math. Phys.
\vol 153 \yr 1993 \pages 159-185
\endref

\ref\key \MMS
\by S.\ Mathur, S.\ Mukhi, A.\ Sen
\paper On the Classification of Rational Conformal Field Theories
\jour Phys. Lett {\bf B}
\vol 213 \yr 1988 \pages 303-308
\endref

\ref\key \Caselle
\by M.\ Caselle, G.\ Ponzano, F.\ Ravanini
\paper Towards a Classification of Fusion Rule Algebras in Rational
       Conformal Field Theories
\jour Int. J. Mod. Phys. {\bf B}
\vol 6 \yr 1992 \pages 2075-2090
\endref

\ref\key \Ehof
\by W.\ Eholzer
\paper Fusion Algebras Induced by Representations of the
       Modular Group
\jour  Int. J. Mod. Phys. {\bf A}
\vol 8 \yr 1993 \pages 3495-3507
\endref

\ref \key \Nobs
\by A.\ Nobs
\paper Die irreduziblen Darstellungen der Gruppen $SL_2(\Z_p)$
       insbesondere $SL_2(\Z_2)$  I
\jour Comment. Math. Helvetici
\vol 51 \yr 1976 \pages  465-489
\endref

\ref
\by $\quad\ $ A.\ Nobs, J.\ Wolfart
\paper Die irreduziblen Darstellungen der Gruppen $SL_2(\Z_p)$
       insbesondere $SL_2(\Z_2)$  II
\jour Comment. Math. Helvetici
\vol 51 \yr 1976 \pages  491-526
\endref

\ref \key \Fuchs
\by J.\ Fuchs
\paper Fusion Rules in Conformal Field Theory
\jour Fortschr. Phys.
\vol 42 \yr 1994 \pages  1-48
\endref

\ref \key \Duke
\by I.B.\ Frenkel, Y. Zhu
\paper Vertex Operator Algebras Associated to Representations
       of Affine and Virasoro Algebras
\jour Duke Math. J.
\vol 66(1) \yr 1992  \pages 123-168
\endref

\ref\key \AMS
\by I.B.\ Frenkel, Y.\ Huang, J.\ Lepowsky
\book  On Axiomatic Approaches to Vertex Operator Algebras and Modules
\bookinfo Memoirs of the American Mathematical Society, Volume 104, Number 494
\publ American Mathematical Society
\publaddr Providence, Rhode Island
\yr 1993
\endref

\ref \key \MIAU
\by W.\ Eholzer, N.\ -P.\ Skoruppa
\paper Modular Invariance and Uniqueness of Conformal Characters
\jour preprint BONN-TH-94-16, MPI-94-67, Commun. Math. Phys.
\toappear
\endref

\ref\key \huanglep
\by   Y.-Z.\ Huang, J.\ Lepowsky
\paper A theory of tensor products for module categories for a
       vertex operator algebra I, II
\jour preprints, hep-th/9309076, hep-th/9309159
\endref


\ref\key \huanglep
\by   Y.-Z.\ Huang
\paper A theory of tensor products for module categories for a
       vertex operator algebra IV
\jour private communication
\endref



\ref\key \Zhu
\by Y.\ Zhu
\paper  Vertex Operator Algebras, Elliptic Functions, and Modular Forms
\jour Ph.D. thesis, Yale University, 1990
\endref


\ref\key \Verlinde
\by E.\ Verlinde
\paper Fusion Rules and Modular Transformations in 2d Conformal Field Theory
\jour Nucl. Phys. {\bf B}
\vol 300 \yr 1988  \pages 360-376
\endref

\ref\key \Vafa
\by C.\ Vafa
\paper Toward Classification of Conformal Theories
\jour Phys. Lett. {\bf B}
\vol 206 \yr 1988  \pages 421-426
\endref

\ref \key \AndersonMoore
\by G.\ Anderson, G.\ Moore
\paper  Rationality in Conformal Field Theory
\jour Commun. Math. Phys.
\vol 117  \yr 1988   \pages  441-450
\endref

\ref \key \Gunnings
\by R.\ C.\ Gunnings
\book  Lectures on Modular Forms
\publ Princeton University Press
\publaddr Princeton, New Yersey
\yr 1962
\endref

\ref\key \Dorn
\by L.\ Dornhoff
\book Group Representation Theory
\publ Marcel Dekker Inc.
\publaddr New York
\yr 1971
\endref

\ref\key \BPZ
\by A.A.\ Belavin, A.M.\ Polyakov, A.B.\ Zamolodchikov
\paper Infinite Conformal Symmetry in Two-Dimensional
       Quantum Field Theory
\jour Nucl. Phys. {\bf B}
\vol 241 \yr 1984 \pages 333-380
\endref

\ref \key \Wang
\by W.\ Wang
\paper  Rationality of Virasoro Vertex Operator Algebras
\jour Int. Research Notices (in Duke Math. J.)
\vol 7 \yr 1993   \pages  197-211
\endref


\ref \key \Don
\by D.\ Zagier
\paper private communication
\endref


\ref \key \Kir
\by E.\ B.\ Kiritsis
\paper Fuchsian Differential Equations for Characters on the Torus:
       A Classification
\jour Nucl. Phys. {\bf B}
\vol 324  \yr 1989  \pages 475-494
\endref


\ref \key \GP
\by C.\ Batut, D.\ Bernardi, H.\ Cohen, M.\ Olivier
\paper PARI-GP
\publ Universit\'e Bordeaux 1
\publaddr Bordeaux
\yr 1989
\endref



\endRefs
\enddocument